\documentclass[prx,aps,twocolumn,superscriptaddress]{revtex4-2}

\usepackage[pdftex]{graphicx}
\usepackage{amsbsy,amssymb,amsmath,bm,mathtools}

\usepackage{xspace}
\usepackage{bm}% bold math
\usepackage{color}

\usepackage{url}
\usepackage[section]{placeins}
\usepackage[colorlinks=true,linkcolor=blue,citecolor=blue]{hyperref}

\usepackage{enumitem}

\begin{document}

%\preprint{Preprint}

%\title{Phase ordering dynamics of a double-exchange Ising model: \\ Machine learning enabled large-scale kinetic Monte Carlo simulations}

\title{Machine learning force-field model for kinetic Monte Carlo simulations \\ of itinerant Ising magnets}

\author{Alexa Tyberg}
\affiliation{Department of Physics, University of Virginia, Charlottesville, VA 22904, USA}

\author{Yunhao Fan}
\affiliation{Department of Physics, University of Virginia, Charlottesville, VA 22904, USA}

\author{Gia-Wei Chern}
\affiliation{Department of Physics, University of Virginia, Charlottesville, VA 22904, USA}

\date{\today}

\begin{abstract}
We present a scalable machine learning (ML) framework for large-scale kinetic Monte Carlo (kMC) simulations of itinerant electron Ising systems. As the effective interactions between Ising spins in such itinerant magnets are mediated by conducting electrons, the calculation of energy change due to a local spin update requires solving an electronic structure problem. Such repeated electronic structure calculations could be overwhelmingly prohibitive for large systems. Assuming the locality principle, a convolutional neural network (CNN) model is developed to directly predict the effective local field and the corresponding energy change associated with a given spin update based on Ising configuration in a finite neighborhood. As the kernel size of the CNN is fixed at a constant, the model can be directly scalable to kMC simulations of large lattices. Our approach is reminiscent of the ML force-field models widely used in first-principles molecular dynamics simulations. Applying our ML framework to a square-lattice double-exchange Ising model, we uncover unusual coarsening of ferromagnetic domains at low temperatures. Our work highlights the potential of ML methods for large-scale modeling of similar itinerant systems with discrete dynamical variables. 
\end{abstract}

\maketitle

\section{Introduction} 

\label{sec:intro}

Machine learning (ML) and data science methods have become transformative tools in physics research, enabling breakthroughs and advancements across a range of domains~\cite{carrasquilla17,carleo17,vanNieuwenburg17,sarma10,bedolla21,carleo19,karniadakis21,boehnlein22}. For example, by combining data-driven inference with domain-specific constraints, ML-based models are able to predict material properties, such as lattice structures or conductivity,  and accelerate the design of new materials by analyzing high-dimensional datasets. The utilization of ML techniques in computational physics has also reinvigorated the field of multi-scale modeling of complex electron systems.  A fundamental issue in multi-scale simulations is the trade-off between efficiency and accuracy of the numerical methods. An accurate treatment of complex quantum materials often requires time-consuming calculations, which significantly limit the accessible system sizes and time scales. ML models acting as universal approximations for high-dimensional functions~\cite{cybenko89,hornik89,barron93} have significantly improved the efficiency of such complex numerical simulations. 

One of the most prominent examples of ML-enabled large-scale modeling is {\em ab initio} molecular dynamics (MD) simulations that are based on ML force-field models~\cite{behler07,bartok10,li15,shapeev16,botu17,smith17,zhang18,behler16,deringer19,mcgibbon17,suwa19,chmiela17,chmiela18,sauceda20}. Contrary to classical MD simulations with empirical force fields, the atomic forces in quantum MD are computed by integrating out electrons on-the-fly as the atomic trajectories are generated~\cite{marx09}. Over the past decade, various ML models have been developed to emulate the time-consuming first-principles electronic structure calculations, mostly based on the density functional theory (DFT). Since a ML force-field model is essentially a complicated classical force-field model which can be efficiently evaluated, the ML approaches thus combine the best of the two worlds: the high-efficiency of classical MD simulations and the accuracy of quantum electronic structure calculations. 

The modern ML force-field models are also canonical examples of transferrable and scalable computational methods. Fundamentally, as argued by W. Kohn, linear-scaling electronic structure methods are possible mainly because of the locality nature or “nearsightedness” principle~\cite{kohn96,prodan05} of many-electron systems. The locality principle does not require the existence of well-localized Wannier-type wavefunctions, but rather results from the wave-mechanical destructive interference in the many-body system. Indeed, in the pioneering work of Behler and Parrinello~\cite{behler07} and Bart\'ok {\em et al.}~\cite{bartok10}, the locality principle was tacitly assumed in their construction of the NN interatomic potential model. In this approach, the total energy of the system is partitioned as $E = \sum_i \epsilon_i$, where $\epsilon_i$ is called the atomic energy and only depends on the local environment of the $i$-th atom~\cite{behler07,bartok10}. The atomic forces are then obtained from derivatives of the predicted energy:~$\mathbf F_i = -\partial E / \partial \mathbf R_i$, where $\mathbf R_i$ is the atomic position vector. Crucially, the complicated dependence of atomic energy $\epsilon_i$ on its neighborhood is approximated by the ML model, which is trained on the condition that both the predicted individual forces $\mathbf F_i$ as well as the total energy $E$ agree with the quantum calculations.

Motivated by the success of such ML-based quantum MD methods, similar ML force-field models have also been developed to enable multi-scale dynamical modeling of several well-known condensed-matter lattice models~\cite{zhang22,zhang22b,cheng23a,zhang20,zhang21,zhang23,cheng23b}.  Of particular interest is the large-scale Landau-Lifshitz-Gilbert (LLG) dynamics simulations of so-called itinerant electron magnets.  Such metallic magnets are characterized by long-range electron-mediated spin-spin interactions and the emergence of complex non-collinear or non-coplanar magnetic orders. Dynamical modeling of complex textures in itinerant spin systems, however, is a computationally challenging task. This is because the local effective magnetic fields, analogous to forces in molecular dynamics, originate from exchange interactions with itinerant electrons and must be computed quantum mechanically. To solve this computational difficulty, the Behler-Parrinello (BP) ML scheme~\cite{behler07,bartok10} has been generalized to build effective magnetic energy models with the accuracy of quantum calculations for itinerant electron magnets~\cite{zhang20,zhang21,zhang23,cheng23b,brannvall22,novikov22}.  

The ML force-field approaches, however, only apply to systems whose evolution is governed by an equation of motion with driving forces originating from the electron degrees of freedom. For similar hybrid classical-quantum systems with {\em discrete} dynamical classical degrees of freedom, such as Ising or Potts variables, the BP-type methods cannot be directly applied. The time evolution of such systems with discrete dynamical variables is often modeled by kinetic Monte Carlo (kMC) methods~\cite{stoll73,binder74} with Metropolis or Glauber dynamics, instead of differential equations. Central to kMC simulations is the calculation of energy difference $\Delta E$ when attempted updates are made to a local discrete variable. Yet, as the energy calculation requires solving the electron structure problem, similar to both quantum MD and itinerant electron systems, each local update is computationally demanding, thus rendering large-scale kMC simulations impossible. 

In this paper, we propose a linear-scaling ML framework for kMC simulations of discrete classical degrees of freedom coupled to itinerant electrons. Our approach can be viewed as the discrete analog of the BP-type ML force-field models. Central to our approach is a deep-learning neural network model which is trained to predict the energy change $\Delta E$ caused by a local update of the discrete variable. Importantly, this energy prediction is assumed to depend only on configurations of the classical variables within a finite neighborhood based on the locality principle. As a result, our proposed ML approach is both transferrable and scalable, which means that the same NN model, successfully trained from small-scale exact solutions, can be applied to much larger systems without rebuilding or retraining.

We demonstrate our approach on an itinerant Ising magnet which can be viewed as a simplified version of the double-exchange (DE) model with strong easy-axis anisotropy. The DE mechanism~\cite{zener51,degennes60,anderson55}, which describes the interplay between ferromagnetism and electron conduction, plays a central role in the emergence of colossal magnetoresistance (CMR) observed in several manganites~\cite{dagotto-book,dagotto01,uehara99,yunoki98,dagotto98}. The Ising-DE model on a square lattice exhibits a ferromagnetic ordered state due to the DE mechanism at low temperatures.  Although the ferromagnetic transition is shown to be in the 2D Ising universality class based on previous ED-MC simulations~\cite{motome01}, the kinetics of the phase transition has yet to be carefully studied. Our large-scale ML-enabled kMC simulations uncover intriguing temperature-dependent phase ordering dynamics of this Ising-DE system. While the coarsening of Ising domains at higher temperatures (but still below the critical temperature) is consistent with the Allen-Cahn domain-growth law for a non-conserved Ising order, a significantly slower growth of Ising domains at lower temperatures is found to follow an anomalous power law with a much smaller exponent. 

The rest of the paper is organized as follows. We discuss the adiabatic dynamics of the Ising double-exchange model in Sec.~\ref{sec:model}, followed by an explanation of the implementation of the ML method. In Sec.~\ref{sec:benchmark}, the ML model is benchmarked against the results by the ED method for $\Delta E$ prediction. The critical temperature and correlation functions are also calculated to justify our proposed method. The ML model is then applied to the dynamical simulation of the Ising double-exchange model. Detailed characterizations of the domain growth are covered in Sec.~\ref{sec:coarsening}. Finally, we conclude the article in Sec.~\ref{sec:summary} with a summary and outlook.

\section{Model and methods}

\label{sec:model}

\subsection{Ising double-exchange model}

We consider the following double-exchange Hamiltonian with Ising spins~\cite{motome01}
\begin{eqnarray}
    {\mathcal{H}} = -t_{\rm nn} \sum_{\langle ij \rangle}\sum_{\sigma = \uparrow, \downarrow} {c}^\dagger_{i\sigma} c^{\,}_{j\sigma} 
    - J_H \sum_i \sigma_i \left( c^\dagger_{i\uparrow} c^{\,}_{i \uparrow} - c^\dagger_{i\downarrow} c^{\,}_{i \downarrow} \right), \nonumber \\
    \label{eqn:DEmodel}
\end{eqnarray}
where Ising spin $\sigma_i = \pm 1$ represents a local magnetic moment at site-$i$, $c^\dagger_{i\sigma}$ ($c^{\,}_{i\sigma}$) is the creation (annihilation) operator of an electron with spin $\sigma = \uparrow, \downarrow$ at site-$i$, $t_{\rm nn}$ represents the hopping coefficient between nearest neighbors, and $J_H$ denotes the on-site Hund's rule coupling between conduction electrons and local spins. In the limit of large coupling $J_H \to \infty$, electrons are forced to align with the local spins in the low energy section of the model. After projecting to the low-energy sector in this limit, the model simplifies to
\begin{eqnarray}
    \mathcal{H} = -\frac{t_{\rm nn}}{2} \sum_{\langle ij \rangle} (1+ \sigma_i \sigma_j) \tilde{c}^\dagger_i \tilde{c}^{\,}_j,
    \label{eqn:strongHund}
\end{eqnarray}
Here $\tilde{c}^\dagger_i$ and $\tilde{c}_i$ are spinless fermion operators. The above Hamiltonian is essentially a disordered tight-binding model with a nearest-neighbor hopping coefficient $t_{\langle ij \rangle} = 1$ or 0 depending on whether the corresponding spin pair is parallel $\sigma_i \sigma_j = +1$ or antiparallel $\sigma_i \sigma_j = -1$. For an inhomogeneous Ising state, the system is composed of ferromagnetic clusters (parallel spins) separated by antiferromagnetic domain walls (antiparallel spins). Since electrons are forbidden to hop across antiparallel spin-pairs, they are confined within individual ferromagnetic clusters. As electrons can gain kinetic energy through delocalization over the cluster, this quantum effect, also known as DE mechanism, thus drives the growth of ferromagnetic domains and the emergence of ferromagnetic order at low temperatures.

Monte Carlo (MC) methods are the main tools to study both the thermodynamic phases and dynamical phenomena of Ising systems. Central to MC simulations is the calculation of energy difference $\Delta E$ when attempting an update to the Ising spin configuration. Due to the complex and often frustrated electron-mediated spin-spin interactions, the Ising-DE model is not amenable to efficient cluster algorithms for updating spins. And for kMC simulations of thermal quench processes to be discussed below, locality of the fundamental dynamics indicates that the evolution of the system is driven by single-spin updates. For such local updates, it is convenient to relate the energy change $\Delta E_i$ due to flipping of $\sigma_i$ to a local effective magnetic field $h_i$:
\begin{eqnarray}
	\label{eq:h_eff}
	\Delta E_i = 2 \sigma_i \, h_i.
\end{eqnarray}
Noting that a spin flip results in a change of magnetic moment $\Delta m_i = (-\sigma_i) - \sigma_i = - 2 \sigma_i$, the local field can be viewed as an effective ``force" given by the discrete analog of the energy derivative: $h_i = -\Delta E_i / \Delta m_i$. 

For short-range Ising models, the calculation of local field $h_i$ is straightforward and can be done rather efficiently; computation of energy difference for itinerant Ising magnets, however, is time-consuming. This is because, even though the tight-binding Hamiltonian is modified only locally by a single-spin flip, the resultant energy change could be rather non-local due to the quantum-mechanical delocalization of electrons. Within the adiabatic approximation, which assumes fast electron relaxation in between spin updates, one can in principle introduce an effective spin-spin interaction of the Ising-DE model in Eq.~(\ref{eqn:strongHund}) by integrating out the electrons. Taking into account the time-reversal symmetry of the model, this effective interaction can be formally expressed as
\begin{eqnarray}
	\label{eq:E_eff}
	& & \mathcal{E} = \sum_{ij} J_{ij} \sigma_i \sigma_j + \sum_{ijkl} K_{ijkl} \sigma_i \sigma_j \sigma_k \sigma_l \nonumber \\
	& & \qquad + \sum_{ijklmn} L_{ijklmn} \sigma_i \sigma_j \sigma_k \sigma_l \sigma_m \sigma_n + \cdots .
\end{eqnarray}
The various terms represent the two-, four-, and six-spin interactions, and so on. The coupling coefficients $J, K$, and $L$ are constrained by the lattice symmetries. Given the above classical spin Hamiltonian, one immediately obtains the local field:
\begin{eqnarray}
	\label{eq:local_h}
	h_i = -\sum_j J_{ij} \sigma_j - \sum_{jkl} K_{jkl} \sigma_j \sigma_k \sigma_l - \cdots .
\end{eqnarray}
For small Hund's coupling $J \ll t_{\rm nn}$, one can perform a many-body perturbation calculation to derive these interaction coefficients. For example, the two-spin coefficient of the order $J_{ij} \sim \mathcal{O}(J^2/t_{\rm nn})$, which is similar to the RKKY interaction for the weak-coupling s-d model~\cite{Ruderman1954,Kasuya1956,Yosida1957}, is expected to decay with the distance $r_{ij}$. For intermediate or large Hund's coupling, a systematic calculation of the multi-spin interactions is extremely tedious, if not impossible. 

On the other hand, since the Ising-DE Hamiltonian in Eq.~(\ref{eqn:strongHund}) is quadratic in electron operators, the energy difference $\Delta E_i$ can be exactly calculated by numerically solving the electron tight-binding Hamiltonian. Specifically, within the adiabatic approximation, the energy change is given by $\Delta E_i = \langle \mathcal{H}(S') \rangle - \langle \mathcal{H}(S) \rangle$, where the two spin configurations $S$ and $S'$ differ by a single-spin flip $\sigma_i \to -\sigma_i$ and each calculation of the expectation value requires solving the corresponding tight-binding Hamiltonian.  
A standard method to solve the tight-binding Hamiltonian is exact diagonalization (ED). For example, from the eigen-solutions, the total system energy at zero temperature is simply $\langle \mathcal{H} \rangle = \sum_{m = 1}^{N_e} \varepsilon_m$, where $\varepsilon_m$ denotes the $m$-th eigen-energy (arranged in ascending order) and $N_e = f N$ is the total number of electrons determined from the filling fraction $f$. However, due to the poor cubic scaling $\mathcal{O}(N^3)$ of ED, the method cannot be used for large-scale MC simulations. 

The kernel polynomial method (KPM) offers a more efficient linear-scaling approach to solve the electron Hamiltonian~\cite{weisse06,barros13,wang18}. In this approach, the total energy is expressed as the integral $\langle \mathcal{H} \rangle = \int^{\varepsilon_F} \varepsilon \,\rho(\varepsilon) d\varepsilon$, where $\varepsilon_F$ is the Fermi level determined by the filling fraction and $\rho(\varepsilon)$ is the density of state (DOS) function. By expanding the DOS in terms of Chebyshev polynomials, the expansion coefficients can be efficiently computed from sparse-matrix-vector multiplications. When combined with MD or LLG simulations, one single electronic structure calculation with KPM is used to update the whole system at a time step, thus realizing linear scalability. However, in MC simulations, every single-spin update requires one KPM calculation, which means the time complexity for one sweep over the system, the basic time unit in MC simulations, scales as $\mathcal{O}(N^2)$. As a result, system sizes that can be feasibly simulated by KPM-based MC are restricted to at most a few hundred~\cite{motome01}.

\begin{figure*}
\centering
\includegraphics[width=1.99\columnwidth]{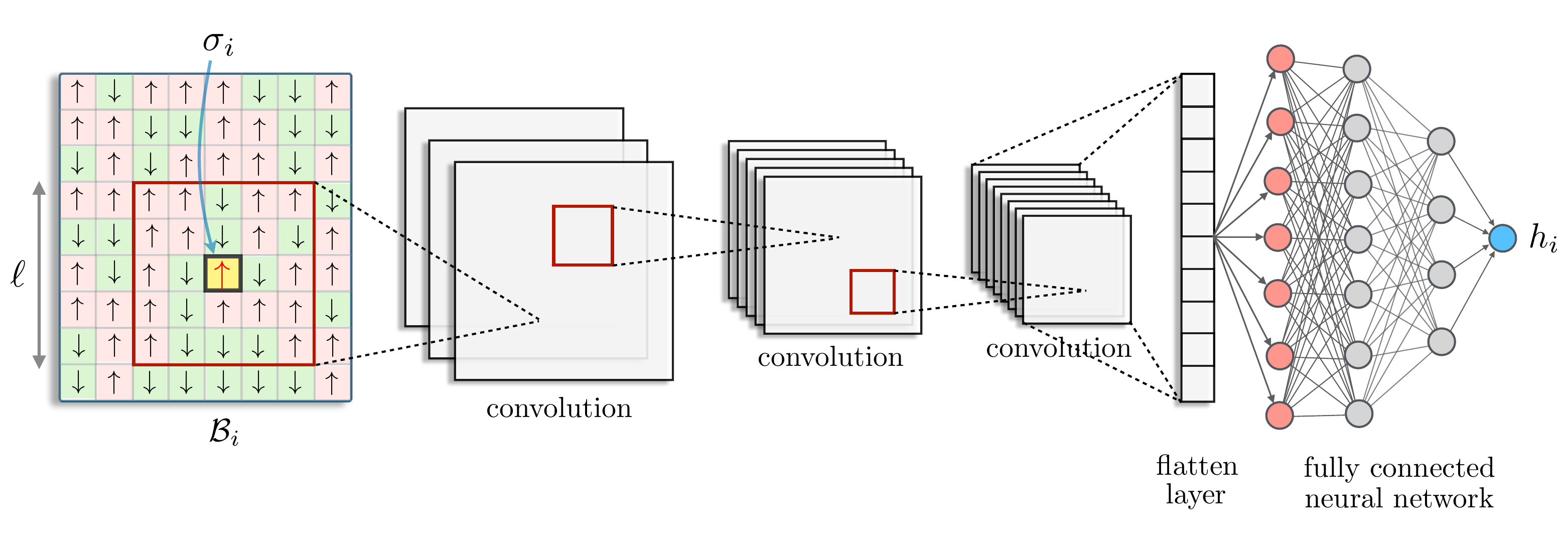}
\caption{Schematic diagram of ML model for prediction of a local effective field $h_i$ associated with a spin $\sigma_i$ on the lattice. The ML model consists of a convolutional neural network (CNN) and a fully connected network. The input to the CNN is Ising configurations $\sigma_j$ within a finite block $\mathcal{B}_i$ in the neighborhood of site-$i$. The single node at the output layer of the fully connected neural net gives the local field $h_i$. The energy change caused by the local spin flip $\sigma_i \to -\sigma_i$ is given by $\Delta E_i = 2 \sigma_i h_i$. }
    \label{fig:cnn}
\end{figure*}

\subsection{ML model for local effective field}

Here we present a linear-scaling method for kMC simulations of itinerant Ising systems based on ML force-field models. As discussed above, linear scalability of computational methods fundamentally relies on the locality principle, and a ML-based divide-and-conquer approach provides a most natural way to take advantage of this locality property for implementing linear-scaling algorithms. Specifically, we assume that the local effective field $h_i$ at a given lattice site-$i$ depends only on a finite neighborhood:
\begin{eqnarray}
	\label{eq:h_ML}
	h_i = \mathcal{F}\bigl( \{ \sigma_j \, \big| \, j \in \mathcal{B}_i \}),
\end{eqnarray}
where $\mathcal{B}_i$ denotes a finite region of linear size $\ell$ centered at site-$i$, and $\mathcal{F}(\cdot)$ is a universal function depending on the electron models being considered. The complex dependence of $h_i$ on the neighborhood Ising configurations is to be learned by a NN model. Specifically, we employ a convolutional neural network (CNN) combined with several fully connected layers of neurons to implement this complex high-dimensional function $\mathcal{F}(\cdot)$. The scale of the NN naturally depends on the size $\ell$ of the neighborhood region, which represents the range of the electron-mediated spin-spin interactions. The spin-flip induced energy change $\Delta E_i$, which is central to MC simulations, is given by Eq.~(\ref{eq:h_eff}). Importantly, since the size of the NN is fixed, the run time of the ML model is independent of the system size. The time complexity of one MC sweep over the system is thus of $\mathcal{O}(N)$.

%Explanation of CNN

A crucial component of our ML model is the CNN, which represents a deep learning algorithm specialized for processing data with a grid-like topology. The convolutional layers in a CNN apply kernels, also known as filters, to perform the mathematical convolution operation on a multidimensional array of input data~\cite{Goodfellow16-ch9}. The filter slides over the input data, and at each sliding window, a dot product of the filter and the corresponding patch of the input is calculated.  The output of the convolution operation with filters is called a feature map, or activation map. A convolutional layer can have multiple filters, each learning a different feature. The output of all filters is stacked along a new dimension, resulting in a multi-channel feature map. The architecture of multiple convolution layers in a CNN makes it particularly suited for exploiting spatial hierarchies in the input data: early layers in a CNN learn simple features like edges or corners, while deeper layers combine these simple features into more complex structures.

A schematic diagram of the neural network model for predicting $\Delta E_i$ is shown in FIG.~\ref{fig:cnn}. The input is an Ising configuration within a $21 \times 21$ square block $\mathcal{B}_i$, where the central spin $\sigma_i$ is the site of the update attempt. After the input layer, 7 convolutional layers are applied with a kernel size of $7 \times 7$ for the first layer and $3 \times 3$ for the rest of the layers.  Though many convolutional neural networks include some pooling or batch normalization layers, the performance of this model is best without either. The number of filters in the respective layers are 6, 12, 16, 20, 24, 24, 24. Following the convolutional layers is a flatten layer to transform the 2D output of the convolutional layers into a 1D array that can be fed into a fully connected network. The fully connected network consists of one layer with 64 nodes followed by an output layer with a single node, which outputs the $\Delta E_i$ energy difference of flipping the spin in the center of the neighborhood.

The Ising-DE models are characterized by both the $Z_2$ time-reversal symmetry as well as the $D_4$ point group symmetry of the square lattice. To incorporate both symmetries into the CNN model, we introduced data augmentation during our training phase.
Data augmentation is a technique commonly used in deep learning to introduce more variety in the data and increase the size of the training dataset by modifying copies of the original data. The increased size of the dataset and the model's exposure to variations on the input make the model more robust against variations which are not supposed to change the output.  In our case, the neighborhood Ising configurations  that are related to each other by a symmetry transformation of the $Z_2 \times D_4$ group are expected to produce the same output $\Delta E_i$. We then expand the dataset by generating 16 symmetry-related configurations for each of the original data entry, all of them associated with exactly the same output $\Delta E_i$.  Through exposure to the expanded dataset, the symmetry properties of the Ising-DE model can then be learned by the CNN model statistically. A similar data-augmentation technique is also used in a recent work to incorporate continuous symmetries in a CNN-based force-field model for itinerant Heisenberg spin systems~\cite{cheng23b}.

Finally, we note that the locality principle means that the interaction coefficients of the spin Hamiltonian Eq.~(\ref{eq:E_eff}), $J_{ij}$, $K_{ijkl}$, $\cdots$, generally decays with increasing separations between spins. This is indeed the case with RKKY interaction, although the decay is a slower algebraic one. But the perturbative derivation of RKKY assumes a gapless electron gas that is unperturbed by the presence of spins~\cite{Ruderman1954,Kasuya1956,Yosida1957}. For intermediate and large spin-electron couplings, the feedback from spins often lead to either an opening of energy gap or localization of electron wave functions. The resultant spin-spin coupling coefficients most likely will decay exponentially. However, analytical calculations of these coefficients in the large $J_H$ regime are generally not possible. the ML force-field model offers a more systematic and efficient approach to obtain the classical effective local field which is formally expressed as a multi-spin expansion in Eq.~(\ref{eq:local_h}).

\section{Benchmarks of ML model}
\label{sec:benchmark}

\begin{figure}[b]
    \centering
    \includegraphics[width=\columnwidth]{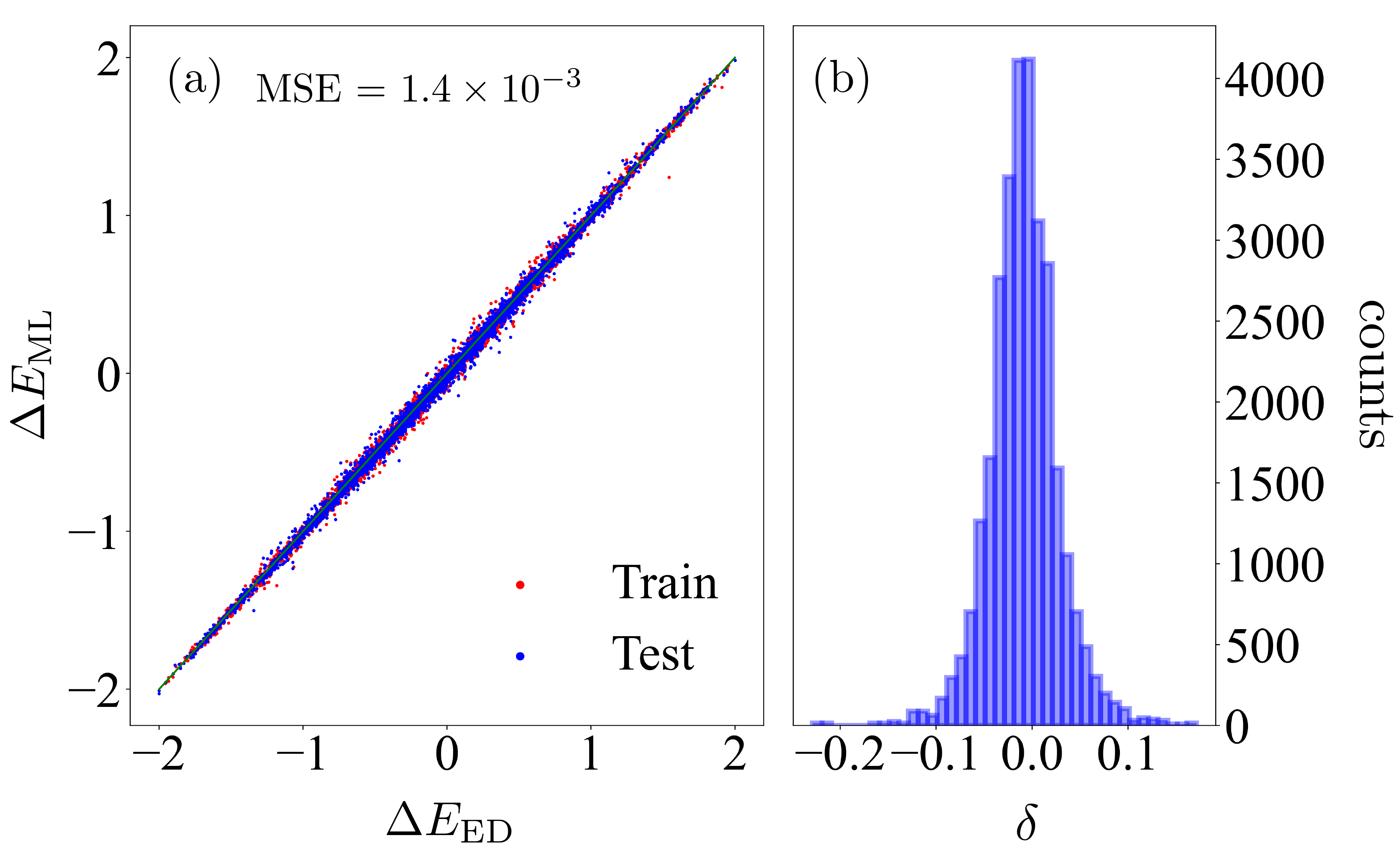}
    \caption{Benchmark of the ML prediction for the energy difference $\Delta E$. Panel (a) compares the ML-predicted energy difference $\Delta E_{\rm ML}$ to the ground-truth value $\Delta E_{\rm ED}$ calculated by exact diagonalization. The mean squared error on the test set is 0.0014. Panel (b) shows a histogram of the prediction error $\delta = \Delta E_{\rm ML} - \Delta E_{\rm ED}$ on the test dataset.}
    \label{fig:pred_benchmark}
\end{figure}

The training and test datasets were generated by ED solutions on a $30 \times 30$ lattice. A total of 100 independent Ising configuration were collected for the full dataset. For each snapshot, each lattice point contributes one entry to the dataset. Specifically, for a given site-$i$, the input is spin configuration within the neighborhood $\mathcal{N}_i$, while the output is the energy change $\Delta_i$ due to flipping of $\sigma_i$. This gives a total of $100 \times 900$ data entries before data augmentation. We then implement data augmentation by applying symmetries in the $Z_2 \times D_4$ group to the neighborhood associated with each $\Delta E_i$ value. This gives a total of $100 \times 900 \times 16$ $\Delta E_i$-neighborhood pairs in the dataset. This was split into train and test sets, with test size approximately 15\% of the total dataset. 

To benchmark the NN model, the $\Delta E$ values predicted by the model are compared with the ground-truth $\Delta E$ values obtained using exact diagonalization for data both in the train and test datasets; see FIG. \ref{fig:pred_benchmark}(a). The predictions overall agree very well with the exact results. We achieve a MSE of 0.0014 on the test set and 0.0013 on the training set. In addition, the training and test sets produce similar error distributions, indicating that there is no overfitting in our ML model. We plot the distribution of the prediction error on the test set in FIG. \ref{fig:pred_benchmark}(b) and see that it follows a Gaussian shape with a standard deviation  of~0.036.

Next we use the trained ML model to examine the thermodynamic behaviors of the Ising-DE model, especially the ferromagnetic phase transition. To this end, we perform Markov Chain Monte Carlo (MCMC) simulations on a $30\times 30$ system using the trained ML model to compute the $\Delta E$. The standard Metropolis algorithm $p_{\rm acc} = \min\{ 1, \exp(-\Delta E / T) \}$ is used to determine the acceptance probability of the attempted spin flip~\cite{Landau_book14}. Here $T$ is the temperature which is measured in units of the nearest-neighbor hopping $t_{\rm nn}$ in the following. After throwing away the 20,000 initial thermalization sweeps to allow the system reach equilibrium, we sampled 500 Ising configurations with 20 sweeps in between the snapshots to compute the thermodynamic properties. Here one sweep is defined as a sequential scan over the entire lattice, applying Metropolis update to each spin along the way. 

\begin{figure}
    \centering
    \includegraphics[width=\columnwidth]{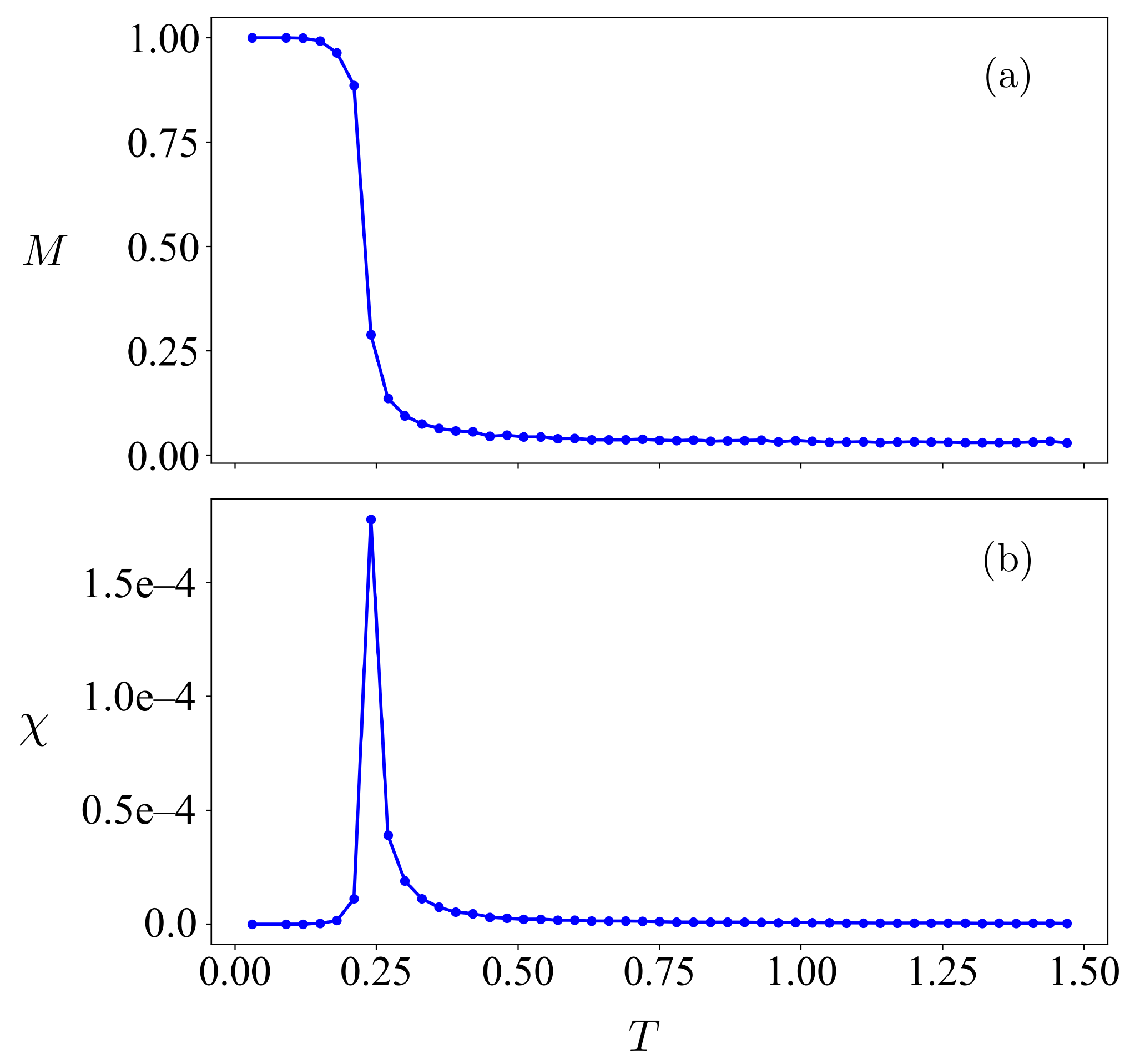}
    \caption{MCMC simulation results based on the ML energy model for the Ising-DE system. (a) magnetization $M$ and (b) magnetic susceptibility $\chi$ versus temperature. The system size is $30 \times 30$ and the temperature is measured in unit of the nearest-neighbor hopping $t_{\rm nn}$. }
    \label{fig:equil}
\end{figure}

In particular, we focus on the temperature dependence of the magnetization $M$, which is the order parameter of the ferromagnetism, and the corresponding susceptibility $\chi$. The ensemble-averaged magnetization of a given snapshot is defined as
\begin{eqnarray}
	M = \frac{1}{N} \Bigl\langle \Bigl| \sum_i \sigma_i \Bigr| \Bigr\rangle,
\end{eqnarray}
where $\langle \cdots \rangle$ indicates average over configurations sampled from MCMC simulations and $N = L^2$ is the total number of spins in a square lattice of linear size $L$. The susceptibility is given by
\begin{eqnarray}
	\chi = \frac{1}{N T} \biggl( \Bigl\langle \Bigl| \sum_i \sigma_i \Bigr|^2 \Bigr\rangle - \Bigl\langle \Bigl| \sum_i \sigma_i \Bigr| \Bigr\rangle^2 \biggr).
\end{eqnarray}
The MCMC results summarized in FIG.~\ref{fig:equil} clearly show a phase transition at $T_c \approx 0.24$. Below the transition temperature, the magnetization $M$ rises to its saturation value $M_{\rm max} = 1$, while the susceptibility exhibits a pronounced peak at $T_c$. The transition temperature obtained from previous MCMC simulations based on a moment-expansion method similar to KPM is $T_c \approx 0.058 W \approx 0.232$, where $W = 4 t_{\rm nn}$ is half the electron bandwidth of the square-lattice tight-binding model~\cite{motome01}. This previous result is in remarkable agreement with the critical temperature obtained from our ML-based MCMC simulations.

\begin{figure}
    \centering
    \includegraphics[width=\columnwidth]{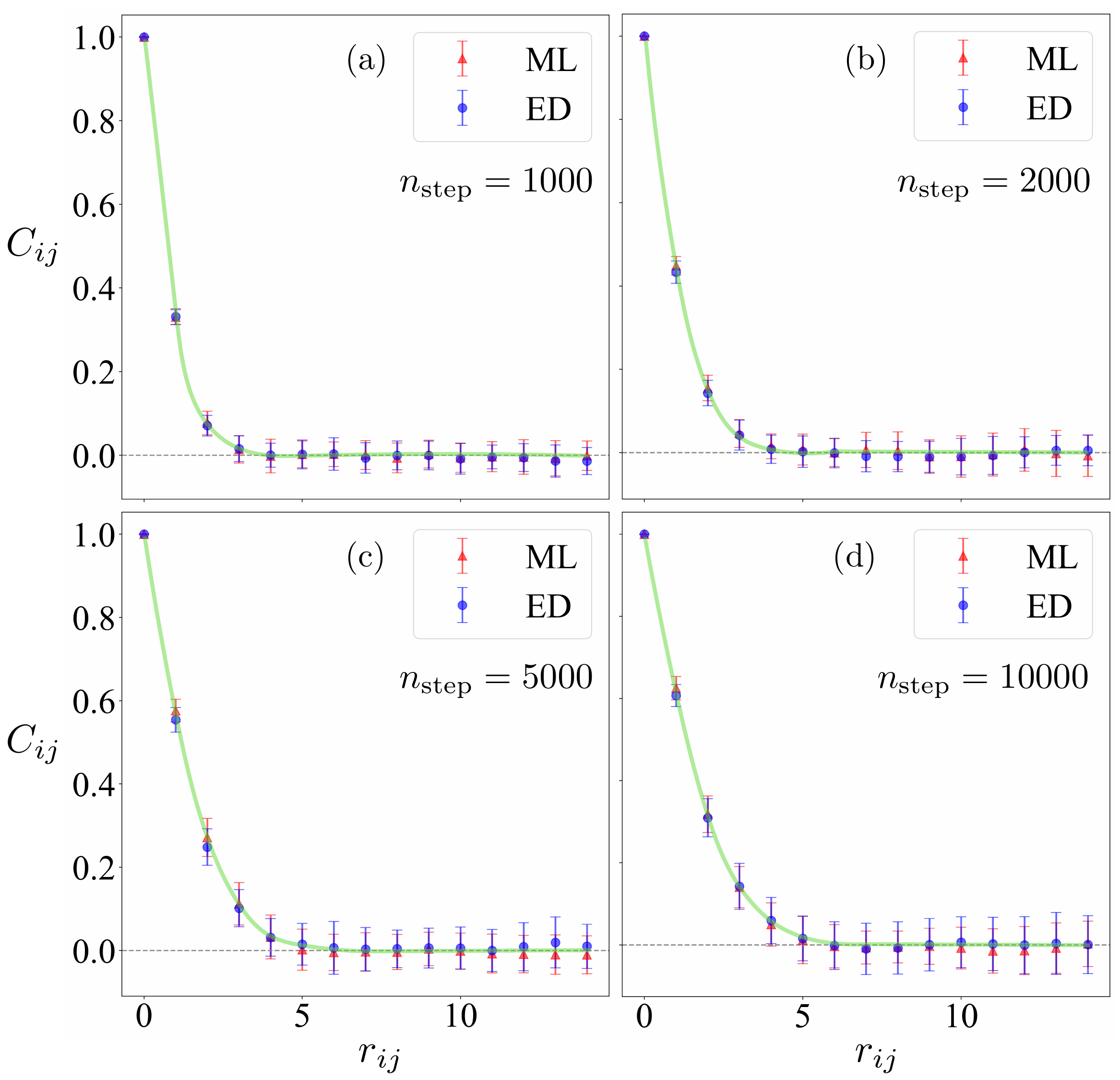}
    \caption{Dynamical benchmark comparison of ML and ED correlation functions at various times after a quench to temperature $T = 0.01$ for a $30 \times 30$ lattice. $n_{step}$ refers to the number of kinetic Monte Carlo spin-update attempts performed before calculating the correlation function. Correlation functions are averaged over 30 independent runs and error bars of $\pm 1$ standard deviation are shown.}
    \label{fig:dynamical_benchmark}
\end{figure}

Since the main interest of this work is on the dynamical evolution of the Ising-DE model, we next perform a dynamical benchmark of the ML energy model. To this end, we carry out kMC simulations of a thermal quench scenario using both ED and ML calculations for the energy change $\Delta E$. A relatively small system of $30\times 30$ lattice was simulated in order to perform the ED-kMC simulations. The kMC algorithm is based on the Glauber dynamics~\cite{Glauber63} for Ising spins, with three major steps: (i)~randomly select a site-$i$ for an attempted spin flip. (ii) Compute the energy difference $\Delta E_i$ due to the spin flip. (iii) Accept the spin flip with a probability $p_i = 1 / (1+e^{- \Delta E_i / T})$, where $T$ is the temperature of the bath after the quench.

For the quench simulations, the system is initially prepared in a state of random spins, corresponding to an equilibrium at infinite temperature. The thermal bath is suddenly quenched to a low-temperature $T = 0.01$ at time $t = 0$. As this temperature is well below the critical temperature of the ferromagnetic phase transition, the subsequent relaxation of the system is dominated by the development of long-range magnetic order. To quantify the development of the ferromagnetic order, we consider the following spin-spin correlation function
\begin{eqnarray}
	\label{eq:C_ij}
	C_{ij}(t) = \langle \sigma_i(t) \sigma_j(t) \rangle - \langle \sigma_i(t) \rangle^2.
\end{eqnarray}
Here $\langle \cdots \rangle$ denotes averaging over spins of a given snapshot as well as ensemble averages (i.e. independent simulations with different initial conditions). FIG.~\ref{fig:dynamical_benchmark} shows the correlation functions, obtained from both ED and ML-kMC simulations, at various times after the thermal quench. Both are obtained by averaging over 30 independent runs.  The correlation functions obtained from the two approaches agree well with each other. This dynamical benchmark indicates that the ML model not only accurately predicts the energy update, but also captures the dynamical evolution of the Ising-DE system.

\section{Coarsening dynamics}

\label{sec:coarsening}

\begin{figure*}
    \centering
    \includegraphics[width=1.99\columnwidth]{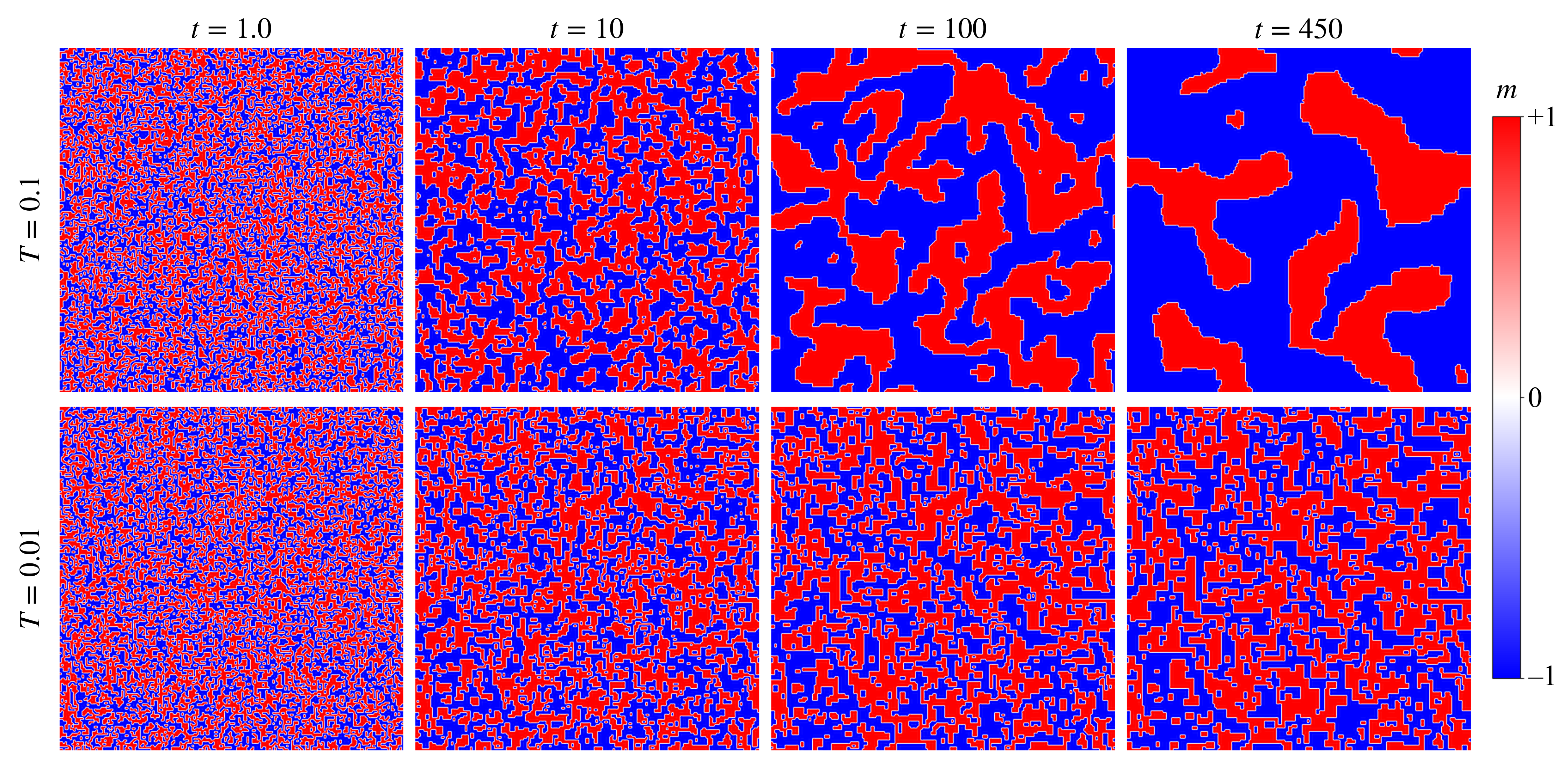}
    \caption{Snapshots of local magnetization $m$ at various times $t$ after a thermal quench of a $200 \times 200$ spin system to temperatures $T=0.1$ and $T=0.01$. The red (blue) regions correspond to ferromagnetic domains of $\sigma_i = +1$ ($\sigma_i = -1$) spins. The system began in a random configuration, and kinetic Monte Carlo simulations with Glauber spin-flip dynamics were paired with the ML $\Delta E$ predictions to simulate its evolution after a sudden temperature quench. $t$ is defined as $n_{\rm step} / N$, where $n_{\rm step}$ is the number of kinetic Monte Carlo spin-update attempts and $N$ is the number of spins in the system.}
    \label{fig:snapshots}
\end{figure*}

On symmetry grounds, the ferromagnetic transition of the Ising-DE model is expected to belong to the 2D Ising universality class. This has been verified through systematic finite-size scaling analysis based on MCMC simulations where the electron structure is solved by KPM~\cite{motome01}. The complex electron-mediated interactions, formally expressed in Eq.~(\ref{eq:E_eff}), seem to produce equilibrium thermodynamic behaviors similar to short-range Ising models. Dynamically, the phase-ordering kinetics of Ising-type symmetry-breaking has also been extensively investigated. For example, it is well established that the coarsening of Ising domains is characterized by a power law $L \sim t^\alpha$, where $L$ is the characteristic domain size and the growth exponent $\alpha = 1/2$ and 1/3 for a nonconserved and conserved, respectively, Ising order parameter~\cite{Bray1994,Onuki2002,Puri2009}. Since the magnetization is not conserved in the Glauber dynamics for updating Ising spins, the growth of the ferromagnetic domains in the Ising-DE model is expected to follow the $\alpha = 1/2$ power law, also known as the Allen-Cahn law.

To study the growth dynamics of ferromagnetic domains, we perform kMC quench simulations with energy difference predicted from our trained ML model. It is worth noting that large-scale simulations are crucial to study the coarsening dynamics since the growth of ordered domains could be affected by finite-size effect for small systems. The ML approach proposed here allows us to carry out quench simulations on lattices with up to $10^5$ spins to properly extract the exponent of the growth power law. As discussed above, the thermal quench scenario corresponds to a sudden change of reservoir temperature from $T = \infty$ to a temperature below the critical point at time $t = 0$. Practically, the system is first initialized in a random state (corresponding to an infinite temperature state). Then Glauber dynamics with $T$ set to the quenched temperature is employed to carry out its time evolution. A scaled simulation time $t = n_{\rm step} / N$  is introduced in order to compare simulation results from different system sizes~\cite{stoll73,binder74}. 

FIG.~\ref{fig:snapshots} shows snapshots of the Ising configuration at different times after a thermal quench for two different quench temperatures $T = 0.1$ and $T = 0.01$. The system size is $200\times 200$. For the case of $T = 0.1$,  the system displays relatively standard coarsening behavior, with overall behaviors similar to the coarsening of ferromagnetic domains in short-range Ising models~\cite{Puri2009, Bray1994}. The phase ordering is characterized by the formation of larger ferromagnetic domains with smooth interfaces. 
By contrast, the evolution in the quench to $T=0.01$ progresses far slower. The domains in this case also have a distinct shape characterized by straight-line boundaries which were not present in the $T=0.1$ case. A qualitative comparison between the two cases indicates temperature-dependent coarsening dynamics of the Ising-DE model, which is in stark contrast to that of standard short-range Ising systems, where coarsening behaviors at different quench temperatures are similar to each other except for the difference in overall time scale.

\begin{figure}
    \centering
    \includegraphics[width=\columnwidth]{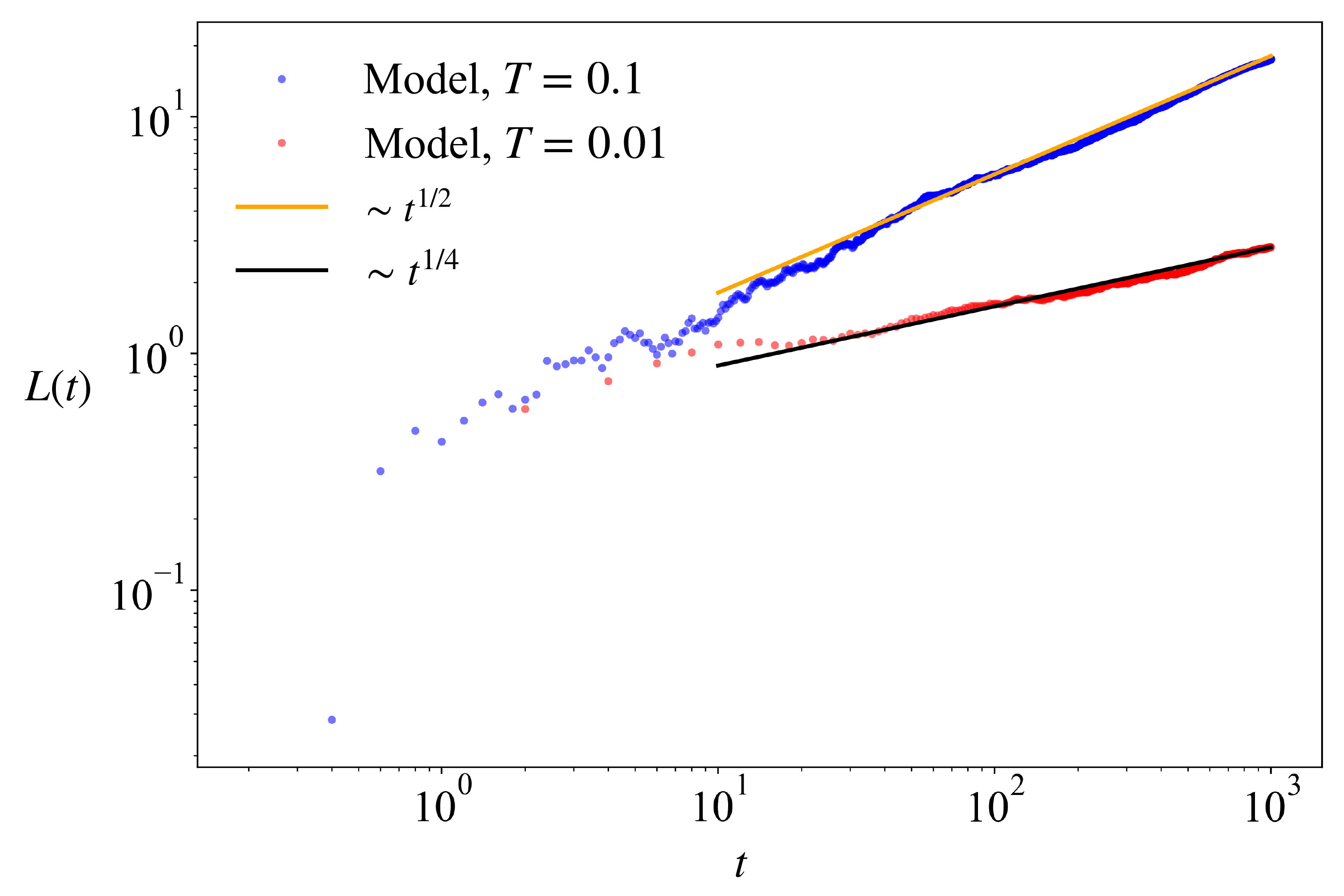}
    \caption{Characteristic domain length $L(t)$ simulated using the ML model on a $100 \times 100$ lattice at temperatures $T=0.1$ and $T=0.01$. At $T=0.1$ the coarsening follows normal Allen-Cahn growth of $L \sim t^{1/2}$ for the standard 2D Ising Model, but at $T=0.01$ the growth rate is reduced to $L \sim t^{1/4}$.}
    \label{fig:Lt}
\end{figure}

To quantify the coarsening dynamics, specifically the growth of ferromagnetic domains, we examine the time-dependent correlation length of the system, which can also be interpreted as the characteristic size of ordered domains. This length scale is computed from the time-dependent correlation functions as follows
\begin{eqnarray}
    L(t) = \frac{\sum_{r} r C(r,t)}{\sum_{r} C(r,t)}
    \label{eqn:L(t)}
\end{eqnarray}
where $C(r, t) = C_{ij}$ is defined in Eq.~(\ref{eq:C_ij}) with the distance $r = r_{ij}$.  We perform a kMC quench simulation on an initially random $100 \times 100$ lattice and calculate $L(t)$ after every 2000 MC steps. For each of the selected time steps, the correlation function is computed by averaging over 20 independent, randomly initialized runs. The time-dependent characteristic length $L(t)$ obtained from the averaged correlation functions is shown in FIG.~\ref{fig:Lt} for the two quench temperatures $T = 0.1$ and $T = 0.01$. The relatively straight lines in the log-log plot indicate that the domain growth for both temperatures exhibits a power-law behavior
\begin{eqnarray}
	\label{eq:Lt_alpha}
	L(t) \sim t^\alpha.
\end{eqnarray}
Interestingly, the growth exponent $\alpha$ is different for the two quench temperatures. For the quench to relatively higher temperature $T = 0.1$, the growth of the characteristic $L$ can be well fitted by the Allen-Cahn power law with $\alpha = 1/2$. On the other hand, for quenches to a lower $T = 0.01$, the coarsening is much slower and is best described by a power law with $\alpha = 1/4$. The significant difference in the growth exponent again shows strong temperature-dependent coarsening dynamics of this itinerant Ising system, and is also in stark contrast with the short-range Ising systems which exhibit the same Allen-Cahn domain growth independent of the quench temperature.

The fact that the coarsening at $T = 0.1$ is described by the Allen-Cahn power law indicates a curvature-driven mechanism for the domain growth. This is also consistent with the emergence of relatively smooth domain walls at late times of the phase ordering at $T = 0.1$; see FIG.~\ref{fig:snapshots}. In this regime, the domain-wall motion is governed by the Allen-Cahn equation $v = - c\kappa$, where $v$ is the normal velocity of domain-wall, $\kappa$ is the curvature of the domain-wall, and $c$ is a proportional coefficient depending on the microscopic details~\cite{Allen1972}. Approximating the interfacial velocity by the domain growth rate $v \sim dL/dt$, and the curvature by the inverse domain size $\kappa \sim 1/L$, one obtains a rate equation for the characteristic length $dL/dt \sim -1/L$. This equation can be readily integrated to give the Allen-Cahn power law $L(t) \sim t^{1/2}$.

\begin{figure}
    \centering
    \includegraphics[width=\columnwidth]{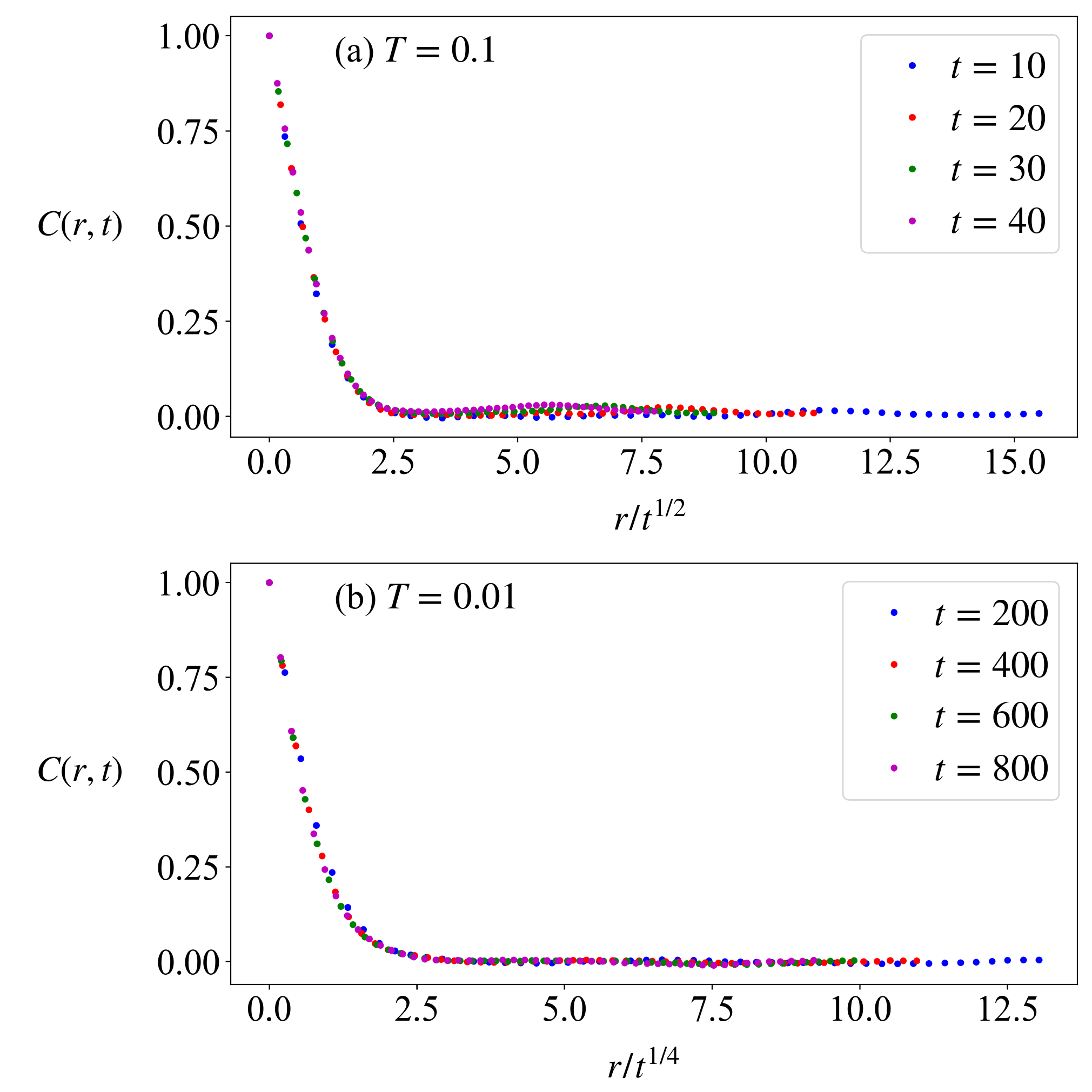}
    \caption{Data point collapse plot of a $100 \times 100$ lattice at temperatures (a) $T=0.1$ and (b) $T=0.01$. The correlation function $C(r,t)$ is graphed against $r/L(t)$, where $L(t) \sim t^{1/2}$ for the higher temperature and $L(t) \sim t^{1/4}$ for the lower temperature.}
    \label{fig:collapse}
\end{figure}

On the other hand, the coarsening of Ising domains at the lower temperature $T = 0.01$ is characterized by a very different morphology. The interfaces are relatively straight and favor either vertical or horizontal directions. Many residual island domains of a few lattice constants remain within larger ordered domains even at late times. This unusual morphology, which is very different from short-range Ising systems, is likely a result of the complex longer-range effective spin-spin interactions, as exemplified by the formal expression in Eq.~(\ref{eq:E_eff}). Importantly, due to the vanishing curvature of a straight domain-wall, the relative abundance of such interfaces would imply a frozen domain-wall motion according to the Allen-Cahn equation~\cite{Allen1972}. The power-law growth with an exponent $\alpha = 1/4$ thus originates from a different domain-growth mechanism, most likely related to a corner-driven scenario. Detailed microscopic mechanisms as well as phenomenological theory of this anomalous coarsening will be left for future studies.

The coarsening behaviors at these two quench temperatures suggest that the domain growth of the Ising-DE model is described by a power-law Eq.~(\ref{eq:Lt_alpha}), yet with a growth exponent which increases monotonically with increasing temperature. Our kMC simulations of quenches at $T \to 0$, based on both ED and ML, found that the system is frozen in disordered states after a very short initial coarsening, suggesting that a vanishing exponent $\alpha = 0$ at zero temperature. The freezing phenomena at $T \to 0$ implies that the system is likely stuck in a local energy minimum. From the viewpoint of the effective spin Hamiltonian Eq.~(\ref{eq:E_eff}), this result suggests a glassy ground state resulting from a highly frustrated spin interaction. Interestingly, a temperature-dependent growth exponent $\alpha(T)$ is also observed in the coarsening dynamics of systems with quenched disorder. The temperature dependence originates from a tunneling through energy barriers assisted by thermal fluctuations~\cite{corberi15}. Yet, since there is no quenched disorder in our simulations of the Ising-DE model, our results thus could be understood from a self-generated disorder through the interplay between the Ising spins and itinerant electrons.

Interestingly, while the domain growth of the Ising-DE system is characterized by the power law of different exponents depending on temperatures, the coarsening of Ising domains at both temperatures exhibits the dynamical scaling property. This means that the domain patterns at different times look statistically similar once scaled globally by the factor $L(t)$ \cite{Bray1994}. Specifically, dynamical scaling suggests that the time-dependence of the system can be fully encoded in the characteristic length $L(t)$, which grows over time. In particular, dynamical scaling indicates the following form for the time-dependent correlation function
\begin{eqnarray}
	C(r, t) = \mathcal{G}\!\left( \frac{r}{L(t)} \right),
\end{eqnarray}
where $\mathcal{G}(x)$ is a universal scaling function of the specific physical system. FIG.~\ref{fig:collapse} shows the correlation function versus the distance $r/L(t)$ rescaled by the time-dependent correlation length. For both quench temperatures, the data points nicely collapse on a hidden curve corresponding to the scaling function, confirming the dynamical scaling symmetry of domain coarsening in both cases of the Ising-DE model.

\section{Summary and Outlook}

\label{sec:summary}

To summarize, we have presented a scalable ML framework for the coarsening dynamics of ferromagnetic domains in the 2D double-exchange model with Ising spins.  We have designed a convolutional neural network, a ML model that is particularly adept at handling multidimensional discrete spin inputs, to predict the local effective field and the corresponding energy change caused by a single-spin flip in itinerant Ising systems. Based on the locality principle, the local field is assumed to depend only on a finite neighborhood of the spin to be updated. As a result, the finite-size NN model, trained by exact diagonalization solutions from small lattices, can be applied to MC simulations on larger systems without rebuilding or retraining, thus ensuring the linear scalability of our approach.   To partially encode the symmetry of the Hamiltonian into the model, data augmentation was used during the model's training. By integrating the ML energy model with Markov Chain and kinetic Monte Carlo simulations, our benchmarks showed that the ML model accurately captures both the equilibrium behavior and the dynamical evolution of the itinerant spin systems.

Large-scale kMC simulations enabled by the ML force-field model uncover unusual phase ordering dynamics of the Ising-DE system. While the coarsening of ferromagnetic domains can be described by a power-law behavior, we find a temperature-dependent growth exponent $\alpha(T)$ which increases from zero in the  $T \to 0$ limit to the Allen-Cahn result $\alpha = 1/2$ at higher quench temperatures. The anomalous temperature-dependent coarsening behaviors are shown to correlate with distinct domain morphologies at different quench temperatures. Despite the unusual domain growth, the coarsening process exhibits dynamical scaling symmetry for all temperatures. 

Unusual domain-coarsening in classical systems is often related to frustrated interactions or quenched disorder~\cite{shore92,evans02,tanaka00,corberi15}. The anomalous phase-ordering of the disorder-free Ising-DE system could be attributed to highly frustrated and long-range effective spin interactions. The temperature-dependent growth exponent, on the other hand, is often attributed to thermal-activated coarsening in systems with quenched disorder. A plausible scenario is the self-generated disorder arising from the interplay between classical Ising spins and quantum electron degrees of freedom.  Given the complexity of such systems, we envision ML techniques as an indispensable tool for multi-scale modeling of nonequilibrium dynamics driven by electron correlation effect. 

Finally, while we demonstrate our approach using the 2D Ising-DE model as an example, the ML framework can be applied to other similar itinerant systems with discrete dynamical degrees of freedom. Fundamentally, spins and other dynamical variables are continuous and governed by differential equations. Discrete degrees of freedom often originates from strong local energy minima, such as easy-axis anisotropy for the case of spins. It is worth noting that discretization in such systems is the first step towards multi-scale dynamical modeling. For discrete systems with more than two fundamental states, such as $q$-state Potts or clock models with $q > 2$, there are more than one possible local transition. As a result, the ML model has to accommodate multiple local effective fields, each corresponds to a local transition and is represented by an output of the neural network. Special care should be taken to properly incorporate both the discrete internal symmetry and the lattice symmetry into the NN model.

\begin{acknowledgments}
The authors thank Sheng Zhang for useful discussions. The work was supported by the US Department of Energy Basic Energy Sciences under Contract No. DE-SC0020330. The authors also acknowledge the support of Research Computing at the University of Virginia.
\end{acknowledgments}

\bibliography{ref}

%apsrev4-2.bst 2019-01-14 (MD) hand-edited version of apsrev4-1.bst
%Control: key (0)
%Control: author (8) initials jnrlst
%Control: editor formatted (1) identically to author
%Control: production of article title (0) allowed
%Control: page (0) single
%Control: year (1) truncated
%Control: production of eprint (0) enabled
\begin{thebibliography}{65}%
\makeatletter
\providecommand \@ifxundefined [1]{%
 \@ifx{#1\undefined}
}%
\providecommand \@ifnum [1]{%
 \ifnum #1\expandafter \@firstoftwo
 \else \expandafter \@secondoftwo
 \fi
}%
\providecommand \@ifx [1]{%
 \ifx #1\expandafter \@firstoftwo
 \else \expandafter \@secondoftwo
 \fi
}%
\providecommand \natexlab [1]{#1}%
\providecommand \enquote  [1]{``#1''}%
\providecommand \bibnamefont  [1]{#1}%
\providecommand \bibfnamefont [1]{#1}%
\providecommand \citenamefont [1]{#1}%
\providecommand \href@noop [0]{\@secondoftwo}%
\providecommand \href [0]{\begingroup \@sanitize@url \@href}%
\providecommand \@href[1]{\@@startlink{#1}\@@href}%
\providecommand \@@href[1]{\endgroup#1\@@endlink}%
\providecommand \@sanitize@url [0]{\catcode `\\12\catcode `\$12\catcode
  `\&12\catcode `\#12\catcode `\^12\catcode `\_12\catcode `\%12\relax}%
\providecommand \@@startlink[1]{}%
\providecommand \@@endlink[0]{}%
\providecommand \url  [0]{\begingroup\@sanitize@url \@url }%
\providecommand \@url [1]{\endgroup\@href {#1}{\urlprefix }}%
\providecommand \urlprefix  [0]{URL }%
\providecommand \Eprint [0]{\href }%
\providecommand \doibase [0]{https://doi.org/}%
\providecommand \selectlanguage [0]{\@gobble}%
\providecommand \bibinfo  [0]{\@secondoftwo}%
\providecommand \bibfield  [0]{\@secondoftwo}%
\providecommand \translation [1]{[#1]}%
\providecommand \BibitemOpen [0]{}%
\providecommand \bibitemStop [0]{}%
\providecommand \bibitemNoStop [0]{.\EOS\space}%
\providecommand \EOS [0]{\spacefactor3000\relax}%
\providecommand \BibitemShut  [1]{\csname bibitem#1\endcsname}%
\let\auto@bib@innerbib\@empty
%</preamble>
\bibitem [{\citenamefont {Carrasquilla}\ and\ \citenamefont
  {Melko}(2017)}]{carrasquilla17}%
  \BibitemOpen
  \bibfield  {author} {\bibinfo {author} {\bibfnamefont {J.}~\bibnamefont
  {Carrasquilla}}\ and\ \bibinfo {author} {\bibfnamefont {R.~G.}\ \bibnamefont
  {Melko}},\ }\bibfield  {title} {\bibinfo {title} {Machine learning phases of
  matter},\ }\href {https://doi.org/10.1038/nphys4035} {\bibfield  {journal}
  {\bibinfo  {journal} {Nature Physics}\ }\textbf {\bibinfo {volume} {13}},\
  \bibinfo {pages} {431} (\bibinfo {year} {2017})}\BibitemShut {NoStop}%
\bibitem [{\citenamefont {Carleo}\ and\ \citenamefont
  {Troyer}(2017)}]{carleo17}%
  \BibitemOpen
  \bibfield  {author} {\bibinfo {author} {\bibfnamefont {G.}~\bibnamefont
  {Carleo}}\ and\ \bibinfo {author} {\bibfnamefont {M.}~\bibnamefont
  {Troyer}},\ }\bibfield  {title} {\bibinfo {title} {Solving the quantum
  many-body problem with artificial neural networks},\ }\href
  {https://doi.org/10.1126/science.aag2302} {\bibfield  {journal} {\bibinfo
  {journal} {Science}\ }\textbf {\bibinfo {volume} {355}},\ \bibinfo {pages}
  {602} (\bibinfo {year} {2017})}\BibitemShut {NoStop}%
\bibitem [{\citenamefont {van Nieuwenburg}\ \emph {et~al.}(2017)\citenamefont
  {van Nieuwenburg}, \citenamefont {Liu},\ and\ \citenamefont
  {Huber}}]{vanNieuwenburg17}%
  \BibitemOpen
  \bibfield  {author} {\bibinfo {author} {\bibfnamefont {E.}~\bibnamefont {van
  Nieuwenburg}}, \bibinfo {author} {\bibfnamefont {Y.-H.}\ \bibnamefont
  {Liu}},\ and\ \bibinfo {author} {\bibfnamefont {S.}~\bibnamefont {Huber}},\
  }\bibfield  {title} {\bibinfo {title} {Learning phase transitions by
  confusion},\ }\href {https://doi.org/10.1038/nphys4037} {\bibfield  {journal}
  {\bibinfo  {journal} {Nature Physics}\ }\textbf {\bibinfo {volume} {13}},\
  \bibinfo {pages} {435} (\bibinfo {year} {2017})}\BibitemShut {NoStop}%
\bibitem [{\citenamefont {Das~Sarma}\ \emph {et~al.}(2019)\citenamefont
  {Das~Sarma}, \citenamefont {Deng},\ and\ \citenamefont {Duan}}]{sarma10}%
  \BibitemOpen
  \bibfield  {author} {\bibinfo {author} {\bibfnamefont {S.}~\bibnamefont
  {Das~Sarma}}, \bibinfo {author} {\bibfnamefont {D.-L.}\ \bibnamefont
  {Deng}},\ and\ \bibinfo {author} {\bibfnamefont {L.-M.}\ \bibnamefont
  {Duan}},\ }\bibfield  {title} {\bibinfo {title} {Machine learning meets
  quantum physics},\ }\href {https://doi.org/10.1063/PT.3.4164} {\bibfield
  {journal} {\bibinfo  {journal} {Physics Today}\ }\textbf {\bibinfo {volume}
  {72}},\ \bibinfo {pages} {48} (\bibinfo {year} {2019})}\BibitemShut {NoStop}%
\bibitem [{\citenamefont {Bedolla}\ \emph {et~al.}(2020)\citenamefont
  {Bedolla}, \citenamefont {Padierna},\ and\ \citenamefont
  {Castañeda-Priego}}]{bedolla21}%
  \BibitemOpen
  \bibfield  {author} {\bibinfo {author} {\bibfnamefont {E.}~\bibnamefont
  {Bedolla}}, \bibinfo {author} {\bibfnamefont {L.~C.}\ \bibnamefont
  {Padierna}},\ and\ \bibinfo {author} {\bibfnamefont {R.}~\bibnamefont
  {Castañeda-Priego}},\ }\bibfield  {title} {\bibinfo {title} {Machine
  learning for condensed matter physics},\ }\href
  {https://doi.org/10.1088/1361-648X/abb895} {\bibfield  {journal} {\bibinfo
  {journal} {Journal of Physics: Condensed Matter}\ }\textbf {\bibinfo {volume}
  {33}},\ \bibinfo {pages} {053001} (\bibinfo {year} {2020})}\BibitemShut
  {NoStop}%
\bibitem [{\citenamefont {Carleo}\ \emph {et~al.}(2019)\citenamefont {Carleo},
  \citenamefont {Cirac}, \citenamefont {Cranmer}, \citenamefont {Daudet},
  \citenamefont {Schuld}, \citenamefont {Tishby}, \citenamefont
  {Vogt-Maranto},\ and\ \citenamefont {Zdeborov\'a}}]{carleo19}%
  \BibitemOpen
  \bibfield  {author} {\bibinfo {author} {\bibfnamefont {G.}~\bibnamefont
  {Carleo}}, \bibinfo {author} {\bibfnamefont {I.}~\bibnamefont {Cirac}},
  \bibinfo {author} {\bibfnamefont {K.}~\bibnamefont {Cranmer}}, \bibinfo
  {author} {\bibfnamefont {L.}~\bibnamefont {Daudet}}, \bibinfo {author}
  {\bibfnamefont {M.}~\bibnamefont {Schuld}}, \bibinfo {author} {\bibfnamefont
  {N.}~\bibnamefont {Tishby}}, \bibinfo {author} {\bibfnamefont
  {L.}~\bibnamefont {Vogt-Maranto}},\ and\ \bibinfo {author} {\bibfnamefont
  {L.}~\bibnamefont {Zdeborov\'a}},\ }\bibfield  {title} {\bibinfo {title}
  {Machine learning and the physical sciences},\ }\href
  {https://doi.org/10.1103/RevModPhys.91.045002} {\bibfield  {journal}
  {\bibinfo  {journal} {Rev. Mod. Phys.}\ }\textbf {\bibinfo {volume} {91}},\
  \bibinfo {pages} {045002} (\bibinfo {year} {2019})}\BibitemShut {NoStop}%
\bibitem [{\citenamefont {Karniadakis}\ \emph {et~al.}(2021)\citenamefont
  {Karniadakis}, \citenamefont {Kevrekidis}, \citenamefont {Lu}, \citenamefont
  {Perdikaris}, \citenamefont {Wang},\ and\ \citenamefont
  {Yang}}]{karniadakis21}%
  \BibitemOpen
  \bibfield  {author} {\bibinfo {author} {\bibfnamefont {G.~E.}\ \bibnamefont
  {Karniadakis}}, \bibinfo {author} {\bibfnamefont {I.~G.}\ \bibnamefont
  {Kevrekidis}}, \bibinfo {author} {\bibfnamefont {L.}~\bibnamefont {Lu}},
  \bibinfo {author} {\bibfnamefont {P.}~\bibnamefont {Perdikaris}}, \bibinfo
  {author} {\bibfnamefont {S.}~\bibnamefont {Wang}},\ and\ \bibinfo {author}
  {\bibfnamefont {L.}~\bibnamefont {Yang}},\ }\bibfield  {title} {\bibinfo
  {title} {Physics-informed machine learning},\ }\href
  {https://doi.org/10.1038/s42254-021-00314-5} {\bibfield  {journal} {\bibinfo
  {journal} {Nature Reviews Physics}\ }\textbf {\bibinfo {volume} {3}},\
  \bibinfo {pages} {422} (\bibinfo {year} {2021})}\BibitemShut {NoStop}%
\bibitem [{\citenamefont {Boehnlein}\ \emph {et~al.}(2022)\citenamefont
  {Boehnlein}, \citenamefont {Diefenthaler}, \citenamefont {Sato},
  \citenamefont {Schram}, \citenamefont {Ziegler}, \citenamefont {Fanelli},
  \citenamefont {Hjorth-Jensen}, \citenamefont {Horn}, \citenamefont {Kuchera},
  \citenamefont {Lee}, \citenamefont {Nazarewicz}, \citenamefont {Ostroumov},
  \citenamefont {Orginos}, \citenamefont {Poon}, \citenamefont {Wang},
  \citenamefont {Scheinker}, \citenamefont {Smith},\ and\ \citenamefont
  {Pang}}]{boehnlein22}%
  \BibitemOpen
  \bibfield  {author} {\bibinfo {author} {\bibfnamefont {A.}~\bibnamefont
  {Boehnlein}}, \bibinfo {author} {\bibfnamefont {M.}~\bibnamefont
  {Diefenthaler}}, \bibinfo {author} {\bibfnamefont {N.}~\bibnamefont {Sato}},
  \bibinfo {author} {\bibfnamefont {M.}~\bibnamefont {Schram}}, \bibinfo
  {author} {\bibfnamefont {V.}~\bibnamefont {Ziegler}}, \bibinfo {author}
  {\bibfnamefont {C.}~\bibnamefont {Fanelli}}, \bibinfo {author} {\bibfnamefont
  {M.}~\bibnamefont {Hjorth-Jensen}}, \bibinfo {author} {\bibfnamefont
  {T.}~\bibnamefont {Horn}}, \bibinfo {author} {\bibfnamefont {M.~P.}\
  \bibnamefont {Kuchera}}, \bibinfo {author} {\bibfnamefont {D.}~\bibnamefont
  {Lee}}, \bibinfo {author} {\bibfnamefont {W.}~\bibnamefont {Nazarewicz}},
  \bibinfo {author} {\bibfnamefont {P.}~\bibnamefont {Ostroumov}}, \bibinfo
  {author} {\bibfnamefont {K.}~\bibnamefont {Orginos}}, \bibinfo {author}
  {\bibfnamefont {A.}~\bibnamefont {Poon}}, \bibinfo {author} {\bibfnamefont
  {X.-N.}\ \bibnamefont {Wang}}, \bibinfo {author} {\bibfnamefont
  {A.}~\bibnamefont {Scheinker}}, \bibinfo {author} {\bibfnamefont {M.~S.}\
  \bibnamefont {Smith}},\ and\ \bibinfo {author} {\bibfnamefont {L.-G.}\
  \bibnamefont {Pang}},\ }\bibfield  {title} {\bibinfo {title} {Colloquium:
  Machine learning in nuclear physics},\ }\href
  {https://doi.org/10.1103/RevModPhys.94.031003} {\bibfield  {journal}
  {\bibinfo  {journal} {Rev. Mod. Phys.}\ }\textbf {\bibinfo {volume} {94}},\
  \bibinfo {pages} {031003} (\bibinfo {year} {2022})}\BibitemShut {NoStop}%
\bibitem [{\citenamefont {Cybenko}(1989)}]{cybenko89}%
  \BibitemOpen
  \bibfield  {author} {\bibinfo {author} {\bibfnamefont {G.}~\bibnamefont
  {Cybenko}},\ }\bibfield  {title} {\bibinfo {title} {Approximation by
  superpositions of a sigmoidal function},\ }\href
  {https://doi.org/10.1007/BF02551274} {\bibfield  {journal} {\bibinfo
  {journal} {Mathematics of Control, Signals and Systems}\ }\textbf {\bibinfo
  {volume} {2}},\ \bibinfo {pages} {303} (\bibinfo {year} {1989})}\BibitemShut
  {NoStop}%
\bibitem [{\citenamefont {Hornik}\ \emph {et~al.}(1989)\citenamefont {Hornik},
  \citenamefont {Stinchcombe},\ and\ \citenamefont {White}}]{hornik89}%
  \BibitemOpen
  \bibfield  {author} {\bibinfo {author} {\bibfnamefont {K.}~\bibnamefont
  {Hornik}}, \bibinfo {author} {\bibfnamefont {M.}~\bibnamefont
  {Stinchcombe}},\ and\ \bibinfo {author} {\bibfnamefont {H.}~\bibnamefont
  {White}},\ }\bibfield  {title} {\bibinfo {title} {Multilayer feedforward
  networks are universal approximators},\ }\href
  {https://doi.org/https://doi.org/10.1016/0893-6080(89)90020-8} {\bibfield
  {journal} {\bibinfo  {journal} {Neural Networks}\ }\textbf {\bibinfo {volume}
  {2}},\ \bibinfo {pages} {359} (\bibinfo {year} {1989})}\BibitemShut {NoStop}%
\bibitem [{\citenamefont {Barron}(1993)}]{barron93}%
  \BibitemOpen
  \bibfield  {author} {\bibinfo {author} {\bibfnamefont {A.}~\bibnamefont
  {Barron}},\ }\bibfield  {title} {\bibinfo {title} {Universal approximation
  bounds for superpositions of a sigmoidal function},\ }\href
  {https://doi.org/10.1109/18.256500} {\bibfield  {journal} {\bibinfo
  {journal} {IEEE Transactions on Information Theory}\ }\textbf {\bibinfo
  {volume} {39}},\ \bibinfo {pages} {930} (\bibinfo {year} {1993})}\BibitemShut
  {NoStop}%
\bibitem [{\citenamefont {Behler}\ and\ \citenamefont
  {Parrinello}(2007)}]{behler07}%
  \BibitemOpen
  \bibfield  {author} {\bibinfo {author} {\bibfnamefont {J.}~\bibnamefont
  {Behler}}\ and\ \bibinfo {author} {\bibfnamefont {M.}~\bibnamefont
  {Parrinello}},\ }\bibfield  {title} {\bibinfo {title} {Generalized
  neural-network representation of high-dimensional potential-energy
  surfaces},\ }\href {https://doi.org/10.1103/PhysRevLett.98.146401} {\bibfield
   {journal} {\bibinfo  {journal} {Phys. Rev. Lett.}\ }\textbf {\bibinfo
  {volume} {98}},\ \bibinfo {pages} {146401} (\bibinfo {year}
  {2007})}\BibitemShut {NoStop}%
\bibitem [{\citenamefont {Bart\'ok}\ \emph {et~al.}(2010)\citenamefont
  {Bart\'ok}, \citenamefont {Payne}, \citenamefont {Kondor},\ and\
  \citenamefont {Cs\'anyi}}]{bartok10}%
  \BibitemOpen
  \bibfield  {author} {\bibinfo {author} {\bibfnamefont {A.~P.}\ \bibnamefont
  {Bart\'ok}}, \bibinfo {author} {\bibfnamefont {M.~C.}\ \bibnamefont {Payne}},
  \bibinfo {author} {\bibfnamefont {R.}~\bibnamefont {Kondor}},\ and\ \bibinfo
  {author} {\bibfnamefont {G.}~\bibnamefont {Cs\'anyi}},\ }\bibfield  {title}
  {\bibinfo {title} {Gaussian approximation potentials: The accuracy of quantum
  mechanics, without the electrons},\ }\href
  {https://doi.org/10.1103/PhysRevLett.104.136403} {\bibfield  {journal}
  {\bibinfo  {journal} {Phys. Rev. Lett.}\ }\textbf {\bibinfo {volume} {104}},\
  \bibinfo {pages} {136403} (\bibinfo {year} {2010})}\BibitemShut {NoStop}%
\bibitem [{\citenamefont {Li}\ \emph {et~al.}(2015)\citenamefont {Li},
  \citenamefont {Kermode},\ and\ \citenamefont {De~Vita}}]{li15}%
  \BibitemOpen
  \bibfield  {author} {\bibinfo {author} {\bibfnamefont {Z.}~\bibnamefont
  {Li}}, \bibinfo {author} {\bibfnamefont {J.~R.}\ \bibnamefont {Kermode}},\
  and\ \bibinfo {author} {\bibfnamefont {A.}~\bibnamefont {De~Vita}},\
  }\bibfield  {title} {\bibinfo {title} {Molecular dynamics with on-the-fly
  machine learning of quantum-mechanical forces},\ }\href
  {https://doi.org/10.1103/PhysRevLett.114.096405} {\bibfield  {journal}
  {\bibinfo  {journal} {Phys. Rev. Lett.}\ }\textbf {\bibinfo {volume} {114}},\
  \bibinfo {pages} {096405} (\bibinfo {year} {2015})}\BibitemShut {NoStop}%
\bibitem [{\citenamefont {Shapeev}(2016)}]{shapeev16}%
  \BibitemOpen
  \bibfield  {author} {\bibinfo {author} {\bibfnamefont {A.~V.}\ \bibnamefont
  {Shapeev}},\ }\bibfield  {title} {\bibinfo {title} {Moment tensor potentials:
  A class of systematically improvable interatomic potentials},\ }\href
  {https://doi.org/10.1137/15M1054183} {\bibfield  {journal} {\bibinfo
  {journal} {Multiscale Modeling \& Simulation}\ }\textbf {\bibinfo {volume}
  {14}},\ \bibinfo {pages} {1153} (\bibinfo {year} {2016})}\BibitemShut
  {NoStop}%
\bibitem [{\citenamefont {Botu}\ \emph {et~al.}(2017)\citenamefont {Botu},
  \citenamefont {Batra}, \citenamefont {Chapman},\ and\ \citenamefont
  {Ramprasad}}]{botu17}%
  \BibitemOpen
  \bibfield  {author} {\bibinfo {author} {\bibfnamefont {V.}~\bibnamefont
  {Botu}}, \bibinfo {author} {\bibfnamefont {R.}~\bibnamefont {Batra}},
  \bibinfo {author} {\bibfnamefont {J.}~\bibnamefont {Chapman}},\ and\ \bibinfo
  {author} {\bibfnamefont {R.}~\bibnamefont {Ramprasad}},\ }\bibfield  {title}
  {\bibinfo {title} {Machine learning force fields: Construction, validation,
  and outlook},\ }\href {https://doi.org/10.1021/acs.jpcc.6b10908} {\bibfield
  {journal} {\bibinfo  {journal} {The Journal of Physical Chemistry C}\
  }\textbf {\bibinfo {volume} {121}},\ \bibinfo {pages} {511} (\bibinfo {year}
  {2017})}\BibitemShut {NoStop}%
\bibitem [{\citenamefont {Smith}\ \emph {et~al.}(2017)\citenamefont {Smith},
  \citenamefont {Isayev},\ and\ \citenamefont {Roitberg}}]{smith17}%
  \BibitemOpen
  \bibfield  {author} {\bibinfo {author} {\bibfnamefont {J.~S.}\ \bibnamefont
  {Smith}}, \bibinfo {author} {\bibfnamefont {O.}~\bibnamefont {Isayev}},\ and\
  \bibinfo {author} {\bibfnamefont {A.~E.}\ \bibnamefont {Roitberg}},\
  }\bibfield  {title} {\bibinfo {title} {Ani-1: an extensible neural network
  potential with dft accuracy at force field computational cost},\ }\href
  {https://doi.org/10.1039/C6SC05720A} {\bibfield  {journal} {\bibinfo
  {journal} {Chem. Sci.}\ }\textbf {\bibinfo {volume} {8}},\ \bibinfo {pages}
  {3192} (\bibinfo {year} {2017})}\BibitemShut {NoStop}%
\bibitem [{\citenamefont {Zhang}\ \emph {et~al.}(2018)\citenamefont {Zhang},
  \citenamefont {Han}, \citenamefont {Wang}, \citenamefont {Car},\ and\
  \citenamefont {E}}]{zhang18}%
  \BibitemOpen
  \bibfield  {author} {\bibinfo {author} {\bibfnamefont {L.}~\bibnamefont
  {Zhang}}, \bibinfo {author} {\bibfnamefont {J.}~\bibnamefont {Han}}, \bibinfo
  {author} {\bibfnamefont {H.}~\bibnamefont {Wang}}, \bibinfo {author}
  {\bibfnamefont {R.}~\bibnamefont {Car}},\ and\ \bibinfo {author}
  {\bibfnamefont {W.}~\bibnamefont {E}},\ }\bibfield  {title} {\bibinfo {title}
  {Deep potential molecular dynamics: A scalable model with the accuracy of
  quantum mechanics},\ }\href {https://doi.org/10.1103/PhysRevLett.120.143001}
  {\bibfield  {journal} {\bibinfo  {journal} {Phys. Rev. Lett.}\ }\textbf
  {\bibinfo {volume} {120}},\ \bibinfo {pages} {143001} (\bibinfo {year}
  {2018})}\BibitemShut {NoStop}%
\bibitem [{\citenamefont {Behler}(2016)}]{behler16}%
  \BibitemOpen
  \bibfield  {author} {\bibinfo {author} {\bibfnamefont {J.}~\bibnamefont
  {Behler}},\ }\bibfield  {title} {\bibinfo {title} {{Perspective: Machine
  learning potentials for atomistic simulations}},\ }\href
  {https://doi.org/10.1063/1.4966192} {\bibfield  {journal} {\bibinfo
  {journal} {The Journal of Chemical Physics}\ }\textbf {\bibinfo {volume}
  {145}},\ \bibinfo {pages} {170901} (\bibinfo {year} {2016})}\BibitemShut
  {NoStop}%
\bibitem [{\citenamefont {Deringer}\ \emph {et~al.}(2019)\citenamefont
  {Deringer}, \citenamefont {Caro},\ and\ \citenamefont {Csanyi}}]{deringer19}%
  \BibitemOpen
  \bibfield  {author} {\bibinfo {author} {\bibfnamefont {V.~L.}\ \bibnamefont
  {Deringer}}, \bibinfo {author} {\bibfnamefont {M.~A.}\ \bibnamefont {Caro}},\
  and\ \bibinfo {author} {\bibfnamefont {G.}~\bibnamefont {Csanyi}},\
  }\bibfield  {title} {\bibinfo {title} {Machine learning interatomic
  potentials as emerging tools for materials science},\ }\href
  {https://doi.org/https://doi.org/10.1002/adma.201902765} {\bibfield
  {journal} {\bibinfo  {journal} {Advanced Materials}\ }\textbf {\bibinfo
  {volume} {31}},\ \bibinfo {pages} {1902765} (\bibinfo {year}
  {2019})}\BibitemShut {NoStop}%
\bibitem [{\citenamefont {McGibbon}\ \emph {et~al.}(2017)\citenamefont
  {McGibbon}, \citenamefont {Taube}, \citenamefont {Donchev}, \citenamefont
  {Siva}, \citenamefont {Hernandez}, \citenamefont {Hargus}, \citenamefont
  {Law}, \citenamefont {Klepeis},\ and\ \citenamefont {Shaw}}]{mcgibbon17}%
  \BibitemOpen
  \bibfield  {author} {\bibinfo {author} {\bibfnamefont {R.~T.}\ \bibnamefont
  {McGibbon}}, \bibinfo {author} {\bibfnamefont {A.~G.}\ \bibnamefont {Taube}},
  \bibinfo {author} {\bibfnamefont {A.~G.}\ \bibnamefont {Donchev}}, \bibinfo
  {author} {\bibfnamefont {K.}~\bibnamefont {Siva}}, \bibinfo {author}
  {\bibfnamefont {F.}~\bibnamefont {Hernandez}}, \bibinfo {author}
  {\bibfnamefont {C.}~\bibnamefont {Hargus}}, \bibinfo {author} {\bibfnamefont
  {K.-H.}\ \bibnamefont {Law}}, \bibinfo {author} {\bibfnamefont {J.~L.}\
  \bibnamefont {Klepeis}},\ and\ \bibinfo {author} {\bibfnamefont {D.~E.}\
  \bibnamefont {Shaw}},\ }\bibfield  {title} {\bibinfo {title} {Improving the
  accuracy of m\"oller-plesset perturbation theory with neural networks},\
  }\href {https://doi.org/10.1063/1.4986081} {\bibfield  {journal} {\bibinfo
  {journal} {The Journal of Chemical Physics}\ }\textbf {\bibinfo {volume}
  {147}},\ \bibinfo {pages} {161725} (\bibinfo {year} {2017})}\BibitemShut
  {NoStop}%
\bibitem [{\citenamefont {Suwa}\ \emph {et~al.}(2019)\citenamefont {Suwa},
  \citenamefont {Smith}, \citenamefont {Lubbers}, \citenamefont {Batista},
  \citenamefont {Chern},\ and\ \citenamefont {Barros}}]{suwa19}%
  \BibitemOpen
  \bibfield  {author} {\bibinfo {author} {\bibfnamefont {H.}~\bibnamefont
  {Suwa}}, \bibinfo {author} {\bibfnamefont {J.~S.}\ \bibnamefont {Smith}},
  \bibinfo {author} {\bibfnamefont {N.}~\bibnamefont {Lubbers}}, \bibinfo
  {author} {\bibfnamefont {C.~D.}\ \bibnamefont {Batista}}, \bibinfo {author}
  {\bibfnamefont {G.-W.}\ \bibnamefont {Chern}},\ and\ \bibinfo {author}
  {\bibfnamefont {K.}~\bibnamefont {Barros}},\ }\bibfield  {title} {\bibinfo
  {title} {Machine learning for molecular dynamics with strongly correlated
  electrons},\ }\href {https://doi.org/10.1103/PhysRevB.99.161107} {\bibfield
  {journal} {\bibinfo  {journal} {Phys. Rev. B}\ }\textbf {\bibinfo {volume}
  {99}},\ \bibinfo {pages} {161107} (\bibinfo {year} {2019})}\BibitemShut
  {NoStop}%
\bibitem [{\citenamefont {Chmiela}\ \emph {et~al.}(2017)\citenamefont
  {Chmiela}, \citenamefont {Tkatchenko}, \citenamefont {Sauceda}, \citenamefont
  {Poltavsky}, \citenamefont {Schütt},\ and\ \citenamefont
  {Müller}}]{chmiela17}%
  \BibitemOpen
  \bibfield  {author} {\bibinfo {author} {\bibfnamefont {S.}~\bibnamefont
  {Chmiela}}, \bibinfo {author} {\bibfnamefont {A.}~\bibnamefont {Tkatchenko}},
  \bibinfo {author} {\bibfnamefont {H.~E.}\ \bibnamefont {Sauceda}}, \bibinfo
  {author} {\bibfnamefont {I.}~\bibnamefont {Poltavsky}}, \bibinfo {author}
  {\bibfnamefont {K.~T.}\ \bibnamefont {Schütt}},\ and\ \bibinfo {author}
  {\bibfnamefont {K.-R.}\ \bibnamefont {Müller}},\ }\bibfield  {title}
  {\bibinfo {title} {Machine learning of accurate energy-conserving molecular
  force fields},\ }\href {https://doi.org/10.1126/sciadv.1603015} {\bibfield
  {journal} {\bibinfo  {journal} {Science Advances}\ }\textbf {\bibinfo
  {volume} {3}},\ \bibinfo {pages} {e1603015} (\bibinfo {year}
  {2017})}\BibitemShut {NoStop}%
\bibitem [{\citenamefont {Chmiela}\ \emph {et~al.}(2018)\citenamefont
  {Chmiela}, \citenamefont {Sauceda}, \citenamefont {M{\"u}ller},\ and\
  \citenamefont {Tkatchenko}}]{chmiela18}%
  \BibitemOpen
  \bibfield  {author} {\bibinfo {author} {\bibfnamefont {S.}~\bibnamefont
  {Chmiela}}, \bibinfo {author} {\bibfnamefont {H.~E.}\ \bibnamefont
  {Sauceda}}, \bibinfo {author} {\bibfnamefont {K.-R.}\ \bibnamefont
  {M{\"u}ller}},\ and\ \bibinfo {author} {\bibfnamefont {A.}~\bibnamefont
  {Tkatchenko}},\ }\bibfield  {title} {\bibinfo {title} {Towards exact
  molecular dynamics simulations with machine-learned force fields},\ }\href
  {https://doi.org/10.1038/s41467-018-06169-2} {\bibfield  {journal} {\bibinfo
  {journal} {Nature Communications}\ }\textbf {\bibinfo {volume} {9}},\
  \bibinfo {pages} {3887} (\bibinfo {year} {2018})}\BibitemShut {NoStop}%
\bibitem [{\citenamefont {Sauceda}\ \emph {et~al.}(2020)\citenamefont
  {Sauceda}, \citenamefont {Gastegger}, \citenamefont {Chmiela}, \citenamefont
  {Müller},\ and\ \citenamefont {Tkatchenko}}]{sauceda20}%
  \BibitemOpen
  \bibfield  {author} {\bibinfo {author} {\bibfnamefont {H.~E.}\ \bibnamefont
  {Sauceda}}, \bibinfo {author} {\bibfnamefont {M.}~\bibnamefont {Gastegger}},
  \bibinfo {author} {\bibfnamefont {S.}~\bibnamefont {Chmiela}}, \bibinfo
  {author} {\bibfnamefont {K.-R.}\ \bibnamefont {Müller}},\ and\ \bibinfo
  {author} {\bibfnamefont {A.}~\bibnamefont {Tkatchenko}},\ }\bibfield  {title}
  {\bibinfo {title} {{Molecular force fields with gradient-domain machine
  learning (GDML): Comparison and synergies with classical force fields}},\
  }\href {https://doi.org/10.1063/5.0023005} {\bibfield  {journal} {\bibinfo
  {journal} {The Journal of Chemical Physics}\ }\textbf {\bibinfo {volume}
  {153}},\ \bibinfo {pages} {124109} (\bibinfo {year} {2020})}\BibitemShut
  {NoStop}%
\bibitem [{\citenamefont {Marx}\ and\ \citenamefont {Hutter}(2009)}]{marx09}%
  \BibitemOpen
  \bibfield  {author} {\bibinfo {author} {\bibfnamefont {D.}~\bibnamefont
  {Marx}}\ and\ \bibinfo {author} {\bibfnamefont {J.}~\bibnamefont {Hutter}},\
  }\href@noop {} {\emph {\bibinfo {title} {Ab initio molecular dynamics: basic
  theory and advanced methods}}}\ (\bibinfo  {publisher} {Cambridge University
  Press},\ \bibinfo {year} {2009})\BibitemShut {NoStop}%
\bibitem [{\citenamefont {Kohn}(1996)}]{kohn96}%
  \BibitemOpen
  \bibfield  {author} {\bibinfo {author} {\bibfnamefont {W.}~\bibnamefont
  {Kohn}},\ }\bibfield  {title} {\bibinfo {title} {Density functional and
  density matrix method scaling linearly with the number of atoms},\ }\href
  {https://doi.org/10.1103/PhysRevLett.76.3168} {\bibfield  {journal} {\bibinfo
   {journal} {Phys. Rev. Lett.}\ }\textbf {\bibinfo {volume} {76}},\ \bibinfo
  {pages} {3168} (\bibinfo {year} {1996})}\BibitemShut {NoStop}%
\bibitem [{\citenamefont {Prodan}\ and\ \citenamefont {Kohn}(2005)}]{prodan05}%
  \BibitemOpen
  \bibfield  {author} {\bibinfo {author} {\bibfnamefont {E.}~\bibnamefont
  {Prodan}}\ and\ \bibinfo {author} {\bibfnamefont {W.}~\bibnamefont {Kohn}},\
  }\bibfield  {title} {\bibinfo {title} {Nearsightedness of electronic
  matter},\ }\href {https://doi.org/10.1073/pnas.0505436102} {\bibfield
  {journal} {\bibinfo  {journal} {Proceedings of the National Academy of
  Sciences}\ }\textbf {\bibinfo {volume} {102}},\ \bibinfo {pages} {11635}
  (\bibinfo {year} {2005})}\BibitemShut {NoStop}%
\bibitem [{\citenamefont {Zhang}\ \emph
  {et~al.}(2022{\natexlab{a}})\citenamefont {Zhang}, \citenamefont {Zhang},\
  and\ \citenamefont {Chern}}]{zhang22}%
  \BibitemOpen
  \bibfield  {author} {\bibinfo {author} {\bibfnamefont {P.}~\bibnamefont
  {Zhang}}, \bibinfo {author} {\bibfnamefont {S.}~\bibnamefont {Zhang}},\ and\
  \bibinfo {author} {\bibfnamefont {G.-W.}\ \bibnamefont {Chern}},\ }\href@noop
  {} {\bibinfo {title} {Descriptors for machine learning model of generalized
  force field in condensed matter systems}} (\bibinfo {year}
  {2022}{\natexlab{a}}),\ \Eprint {https://arxiv.org/abs/2201.00798}
  {arXiv:2201.00798 [cond-mat.str-el]} \BibitemShut {NoStop}%
\bibitem [{\citenamefont {Zhang}\ \emph
  {et~al.}(2022{\natexlab{b}})\citenamefont {Zhang}, \citenamefont {Zhang},\
  and\ \citenamefont {Chern}}]{zhang22b}%
  \BibitemOpen
  \bibfield  {author} {\bibinfo {author} {\bibfnamefont {S.}~\bibnamefont
  {Zhang}}, \bibinfo {author} {\bibfnamefont {P.}~\bibnamefont {Zhang}},\ and\
  \bibinfo {author} {\bibfnamefont {G.-W.}\ \bibnamefont {Chern}},\ }\bibfield
  {title} {\bibinfo {title} {Anomalous phase separation in a correlated
  electron system: Machine-learning enabled large-scale kinetic monte carlo
  simulations},\ }\href {https://doi.org/10.1073/pnas.2119957119} {\bibfield
  {journal} {\bibinfo  {journal} {Proceedings of the National Academy of
  Sciences}\ }\textbf {\bibinfo {volume} {119}},\ \bibinfo {pages}
  {e2119957119} (\bibinfo {year} {2022}{\natexlab{b}})}\BibitemShut {NoStop}%
\bibitem [{\citenamefont {Cheng}\ \emph
  {et~al.}(2023{\natexlab{a}})\citenamefont {Cheng}, \citenamefont {Zhang},\
  and\ \citenamefont {Chern}}]{cheng23a}%
  \BibitemOpen
  \bibfield  {author} {\bibinfo {author} {\bibfnamefont {C.}~\bibnamefont
  {Cheng}}, \bibinfo {author} {\bibfnamefont {S.}~\bibnamefont {Zhang}},\ and\
  \bibinfo {author} {\bibfnamefont {G.-W.}\ \bibnamefont {Chern}},\ }\bibfield
  {title} {\bibinfo {title} {Machine learning for phase ordering dynamics of
  charge density waves},\ }\href {https://doi.org/10.1103/PhysRevB.108.014301}
  {\bibfield  {journal} {\bibinfo  {journal} {Phys. Rev. B}\ }\textbf {\bibinfo
  {volume} {108}},\ \bibinfo {pages} {014301} (\bibinfo {year}
  {2023}{\natexlab{a}})}\BibitemShut {NoStop}%
\bibitem [{\citenamefont {Zhang}\ \emph {et~al.}(2020)\citenamefont {Zhang},
  \citenamefont {Saha},\ and\ \citenamefont {Chern}}]{zhang20}%
  \BibitemOpen
  \bibfield  {author} {\bibinfo {author} {\bibfnamefont {P.}~\bibnamefont
  {Zhang}}, \bibinfo {author} {\bibfnamefont {P.}~\bibnamefont {Saha}},\ and\
  \bibinfo {author} {\bibfnamefont {G.-W.}\ \bibnamefont {Chern}},\ }\href@noop
  {} {\bibinfo {title} {Machine learning dynamics of phase separation in
  correlated electron magnets}} (\bibinfo {year} {2020}),\ \Eprint
  {https://arxiv.org/abs/2006.04205} {arXiv:2006.04205 [cond-mat.str-el]}
  \BibitemShut {NoStop}%
\bibitem [{\citenamefont {Zhang}\ and\ \citenamefont {Chern}(2021)}]{zhang21}%
  \BibitemOpen
  \bibfield  {author} {\bibinfo {author} {\bibfnamefont {P.}~\bibnamefont
  {Zhang}}\ and\ \bibinfo {author} {\bibfnamefont {G.-W.}\ \bibnamefont
  {Chern}},\ }\bibfield  {title} {\bibinfo {title} {Arrested phase separation
  in double-exchange models: Large-scale simulation enabled by machine
  learning},\ }\href {https://doi.org/10.1103/PhysRevLett.127.146401}
  {\bibfield  {journal} {\bibinfo  {journal} {Phys. Rev. Lett.}\ }\textbf
  {\bibinfo {volume} {127}},\ \bibinfo {pages} {146401} (\bibinfo {year}
  {2021})}\BibitemShut {NoStop}%
\bibitem [{\citenamefont {Zhang}\ and\ \citenamefont {Chern}(2023)}]{zhang23}%
  \BibitemOpen
  \bibfield  {author} {\bibinfo {author} {\bibfnamefont {P.}~\bibnamefont
  {Zhang}}\ and\ \bibinfo {author} {\bibfnamefont {G.-W.}\ \bibnamefont
  {Chern}},\ }\bibfield  {title} {\bibinfo {title} {Machine learning
  nonequilibrium electron forces for spin dynamics of itinerant magnets},\
  }\href {https://doi.org/10.1038/s41524-023-00990-0} {\bibfield  {journal}
  {\bibinfo  {journal} {npj Computational Materials}\ }\textbf {\bibinfo
  {volume} {9}},\ \bibinfo {pages} {32} (\bibinfo {year} {2023})}\BibitemShut
  {NoStop}%
\bibitem [{\citenamefont {Cheng}\ \emph
  {et~al.}(2023{\natexlab{b}})\citenamefont {Cheng}, \citenamefont {Zhang},
  \citenamefont {Nguyen}, \citenamefont {Azarfar}, \citenamefont {Chern},\ and\
  \citenamefont {Baek}}]{cheng23b}%
  \BibitemOpen
  \bibfield  {author} {\bibinfo {author} {\bibfnamefont {X.}~\bibnamefont
  {Cheng}}, \bibinfo {author} {\bibfnamefont {S.}~\bibnamefont {Zhang}},
  \bibinfo {author} {\bibfnamefont {P.~C.~H.}\ \bibnamefont {Nguyen}}, \bibinfo
  {author} {\bibfnamefont {S.}~\bibnamefont {Azarfar}}, \bibinfo {author}
  {\bibfnamefont {G.-W.}\ \bibnamefont {Chern}},\ and\ \bibinfo {author}
  {\bibfnamefont {S.~S.}\ \bibnamefont {Baek}},\ }\bibfield  {title} {\bibinfo
  {title} {Convolutional neural networks for large-scale dynamical modeling of
  itinerant magnets},\ }\href
  {https://doi.org/10.1103/PhysRevResearch.5.033188} {\bibfield  {journal}
  {\bibinfo  {journal} {Phys. Rev. Res.}\ }\textbf {\bibinfo {volume} {5}},\
  \bibinfo {pages} {033188} (\bibinfo {year} {2023}{\natexlab{b}})}\BibitemShut
  {NoStop}%
\bibitem [{\citenamefont {Arale~Br\"annvall}\ \emph {et~al.}(2022)\citenamefont
  {Arale~Br\"annvall}, \citenamefont {Gambino}, \citenamefont {Armiento},\ and\
  \citenamefont {Alling}}]{brannvall22}%
  \BibitemOpen
  \bibfield  {author} {\bibinfo {author} {\bibfnamefont {M.}~\bibnamefont
  {Arale~Br\"annvall}}, \bibinfo {author} {\bibfnamefont {D.}~\bibnamefont
  {Gambino}}, \bibinfo {author} {\bibfnamefont {R.}~\bibnamefont {Armiento}},\
  and\ \bibinfo {author} {\bibfnamefont {B.}~\bibnamefont {Alling}},\
  }\bibfield  {title} {\bibinfo {title} {Machine learning approach for
  longitudinal spin fluctuation effects in bcc fe at ${T}_{c}$ and under
  earth-core conditions},\ }\href {https://doi.org/10.1103/PhysRevB.105.144417}
  {\bibfield  {journal} {\bibinfo  {journal} {Phys. Rev. B}\ }\textbf {\bibinfo
  {volume} {105}},\ \bibinfo {pages} {144417} (\bibinfo {year}
  {2022})}\BibitemShut {NoStop}%
\bibitem [{\citenamefont {Novikov}\ \emph {et~al.}(2022)\citenamefont
  {Novikov}, \citenamefont {Grabowski}, \citenamefont {K{\"o}rmann},\ and\
  \citenamefont {Shapeev}}]{novikov22}%
  \BibitemOpen
  \bibfield  {author} {\bibinfo {author} {\bibfnamefont {I.}~\bibnamefont
  {Novikov}}, \bibinfo {author} {\bibfnamefont {B.}~\bibnamefont {Grabowski}},
  \bibinfo {author} {\bibfnamefont {F.}~\bibnamefont {K{\"o}rmann}},\ and\
  \bibinfo {author} {\bibfnamefont {A.}~\bibnamefont {Shapeev}},\ }\bibfield
  {title} {\bibinfo {title} {Magnetic moment tensor potentials for collinear
  spin-polarized materials reproduce different magnetic states of bcc fe},\
  }\href {https://doi.org/10.1038/s41524-022-00696-9} {\bibfield  {journal}
  {\bibinfo  {journal} {npj Computational Materials}\ }\textbf {\bibinfo
  {volume} {8}},\ \bibinfo {pages} {13} (\bibinfo {year} {2022})}\BibitemShut
  {NoStop}%
\bibitem [{\citenamefont {Stoll}\ \emph {et~al.}(1973)\citenamefont {Stoll},
  \citenamefont {Binder},\ and\ \citenamefont {Schneider}}]{stoll73}%
  \BibitemOpen
  \bibfield  {author} {\bibinfo {author} {\bibfnamefont {E.}~\bibnamefont
  {Stoll}}, \bibinfo {author} {\bibfnamefont {K.}~\bibnamefont {Binder}},\ and\
  \bibinfo {author} {\bibfnamefont {T.}~\bibnamefont {Schneider}},\ }\bibfield
  {title} {\bibinfo {title} {Monte carlo investigation of dynamic critical
  phenomena in the two-dimensional kinetic ising model},\ }\href
  {https://doi.org/10.1103/PhysRevB.8.3266} {\bibfield  {journal} {\bibinfo
  {journal} {Phys. Rev. B}\ }\textbf {\bibinfo {volume} {8}},\ \bibinfo {pages}
  {3266} (\bibinfo {year} {1973})}\BibitemShut {NoStop}%
\bibitem [{\citenamefont {Binder}\ and\ \citenamefont
  {M\"uller-Krumbhaar}(1974)}]{binder74}%
  \BibitemOpen
  \bibfield  {author} {\bibinfo {author} {\bibfnamefont {K.}~\bibnamefont
  {Binder}}\ and\ \bibinfo {author} {\bibfnamefont {H.}~\bibnamefont
  {M\"uller-Krumbhaar}},\ }\bibfield  {title} {\bibinfo {title} {Investigation
  of metastable states and nucleation in the kinetic ising model},\ }\href
  {https://doi.org/10.1103/PhysRevB.9.2328} {\bibfield  {journal} {\bibinfo
  {journal} {Phys. Rev. B}\ }\textbf {\bibinfo {volume} {9}},\ \bibinfo {pages}
  {2328} (\bibinfo {year} {1974})}\BibitemShut {NoStop}%
\bibitem [{\citenamefont {Zener}(1951)}]{zener51}%
  \BibitemOpen
  \bibfield  {author} {\bibinfo {author} {\bibfnamefont {C.}~\bibnamefont
  {Zener}},\ }\bibfield  {title} {\bibinfo {title} {Interaction between the
  $d$-shells in the transition metals. ii. ferromagnetic compounds of manganese
  with perovskite structure},\ }\href {https://doi.org/10.1103/PhysRev.82.403}
  {\bibfield  {journal} {\bibinfo  {journal} {Phys. Rev.}\ }\textbf {\bibinfo
  {volume} {82}},\ \bibinfo {pages} {403} (\bibinfo {year} {1951})}\BibitemShut
  {NoStop}%
\bibitem [{\citenamefont {de~Gennes}(1960)}]{degennes60}%
  \BibitemOpen
  \bibfield  {author} {\bibinfo {author} {\bibfnamefont {P.~G.}\ \bibnamefont
  {de~Gennes}},\ }\bibfield  {title} {\bibinfo {title} {Effects of double
  exchange in magnetic crystals},\ }\href
  {https://doi.org/10.1103/PhysRev.118.141} {\bibfield  {journal} {\bibinfo
  {journal} {Phys. Rev.}\ }\textbf {\bibinfo {volume} {118}},\ \bibinfo {pages}
  {141} (\bibinfo {year} {1960})}\BibitemShut {NoStop}%
\bibitem [{\citenamefont {Anderson}\ and\ \citenamefont
  {Hasegawa}(1955)}]{anderson55}%
  \BibitemOpen
  \bibfield  {author} {\bibinfo {author} {\bibfnamefont {P.~W.}\ \bibnamefont
  {Anderson}}\ and\ \bibinfo {author} {\bibfnamefont {H.}~\bibnamefont
  {Hasegawa}},\ }\bibfield  {title} {\bibinfo {title} {Considerations on double
  exchange},\ }\href {https://doi.org/10.1103/PhysRev.100.675} {\bibfield
  {journal} {\bibinfo  {journal} {Phys. Rev.}\ }\textbf {\bibinfo {volume}
  {100}},\ \bibinfo {pages} {675} (\bibinfo {year} {1955})}\BibitemShut
  {NoStop}%
\bibitem [{\citenamefont {Dagotto}(2002)}]{dagotto-book}%
  \BibitemOpen
  \bibfield  {author} {\bibinfo {author} {\bibfnamefont {E.}~\bibnamefont
  {Dagotto}},\ }\href@noop {} {\emph {\bibinfo {title} {Nanoscale Phase
  Separation and Colossal Magnetoresistance}}}\ (\bibinfo  {publisher}
  {Springer Berlin},\ \bibinfo {year} {2002})\BibitemShut {NoStop}%
\bibitem [{\citenamefont {Dagotto}\ \emph {et~al.}(2001)\citenamefont
  {Dagotto}, \citenamefont {Hotta},\ and\ \citenamefont {Moreo}}]{dagotto01}%
  \BibitemOpen
  \bibfield  {author} {\bibinfo {author} {\bibfnamefont {E.}~\bibnamefont
  {Dagotto}}, \bibinfo {author} {\bibfnamefont {T.}~\bibnamefont {Hotta}},\
  and\ \bibinfo {author} {\bibfnamefont {A.}~\bibnamefont {Moreo}},\ }\bibfield
   {title} {\bibinfo {title} {Colossal magnetoresistant materials: the key role
  of phase separation},\ }\href
  {https://doi.org/https://doi.org/10.1016/S0370-1573(00)00121-6} {\bibfield
  {journal} {\bibinfo  {journal} {Physics Reports}\ }\textbf {\bibinfo {volume}
  {344}},\ \bibinfo {pages} {1} (\bibinfo {year} {2001})}\BibitemShut {NoStop}%
\bibitem [{\citenamefont {Uehara}\ \emph {et~al.}(1999)\citenamefont {Uehara},
  \citenamefont {Mori}, \citenamefont {Chen},\ and\ \citenamefont
  {Cheong}}]{uehara99}%
  \BibitemOpen
  \bibfield  {author} {\bibinfo {author} {\bibfnamefont {M.}~\bibnamefont
  {Uehara}}, \bibinfo {author} {\bibfnamefont {S.}~\bibnamefont {Mori}},
  \bibinfo {author} {\bibfnamefont {C.~H.}\ \bibnamefont {Chen}},\ and\
  \bibinfo {author} {\bibfnamefont {S.-W.}\ \bibnamefont {Cheong}},\ }\bibfield
   {title} {\bibinfo {title} {Percolative phase separation underlies colossal
  magnetoresistance in mixed-valent manganites},\ }\href
  {https://doi.org/10.1038/21142} {\bibfield  {journal} {\bibinfo  {journal}
  {Nature}\ }\textbf {\bibinfo {volume} {399}},\ \bibinfo {pages} {560}
  (\bibinfo {year} {1999})}\BibitemShut {NoStop}%
\bibitem [{\citenamefont {Yunoki}\ \emph {et~al.}(1998)\citenamefont {Yunoki},
  \citenamefont {Hu}, \citenamefont {Malvezzi}, \citenamefont {Moreo},
  \citenamefont {Furukawa},\ and\ \citenamefont {Dagotto}}]{yunoki98}%
  \BibitemOpen
  \bibfield  {author} {\bibinfo {author} {\bibfnamefont {S.}~\bibnamefont
  {Yunoki}}, \bibinfo {author} {\bibfnamefont {J.}~\bibnamefont {Hu}}, \bibinfo
  {author} {\bibfnamefont {A.~L.}\ \bibnamefont {Malvezzi}}, \bibinfo {author}
  {\bibfnamefont {A.}~\bibnamefont {Moreo}}, \bibinfo {author} {\bibfnamefont
  {N.}~\bibnamefont {Furukawa}},\ and\ \bibinfo {author} {\bibfnamefont
  {E.}~\bibnamefont {Dagotto}},\ }\bibfield  {title} {\bibinfo {title} {Phase
  separation in electronic models for manganites},\ }\href
  {https://doi.org/10.1103/PhysRevLett.80.845} {\bibfield  {journal} {\bibinfo
  {journal} {Phys. Rev. Lett.}\ }\textbf {\bibinfo {volume} {80}},\ \bibinfo
  {pages} {845} (\bibinfo {year} {1998})}\BibitemShut {NoStop}%
\bibitem [{\citenamefont {Dagotto}\ \emph {et~al.}(1998)\citenamefont
  {Dagotto}, \citenamefont {Yunoki}, \citenamefont {Malvezzi}, \citenamefont
  {Moreo}, \citenamefont {Hu}, \citenamefont {Capponi}, \citenamefont
  {Poilblanc},\ and\ \citenamefont {Furukawa}}]{dagotto98}%
  \BibitemOpen
  \bibfield  {author} {\bibinfo {author} {\bibfnamefont {E.}~\bibnamefont
  {Dagotto}}, \bibinfo {author} {\bibfnamefont {S.}~\bibnamefont {Yunoki}},
  \bibinfo {author} {\bibfnamefont {A.~L.}\ \bibnamefont {Malvezzi}}, \bibinfo
  {author} {\bibfnamefont {A.}~\bibnamefont {Moreo}}, \bibinfo {author}
  {\bibfnamefont {J.}~\bibnamefont {Hu}}, \bibinfo {author} {\bibfnamefont
  {S.}~\bibnamefont {Capponi}}, \bibinfo {author} {\bibfnamefont
  {D.}~\bibnamefont {Poilblanc}},\ and\ \bibinfo {author} {\bibfnamefont
  {N.}~\bibnamefont {Furukawa}},\ }\bibfield  {title} {\bibinfo {title}
  {Ferromagnetic {Kondo} model for manganites: Phase diagram, charge
  segregation, and influence of quantum localized spins},\ }\href
  {https://doi.org/10.1103/PhysRevB.58.6414} {\bibfield  {journal} {\bibinfo
  {journal} {Phys. Rev. B}\ }\textbf {\bibinfo {volume} {58}},\ \bibinfo
  {pages} {6414} (\bibinfo {year} {1998})}\BibitemShut {NoStop}%
\bibitem [{\citenamefont {Motome}\ and\ \citenamefont
  {Furukawa}(2001)}]{motome01}%
  \BibitemOpen
  \bibfield  {author} {\bibinfo {author} {\bibfnamefont {Y.}~\bibnamefont
  {Motome}}\ and\ \bibinfo {author} {\bibfnamefont {N.}~\bibnamefont
  {Furukawa}},\ }\bibfield  {title} {\bibinfo {title} {Critical phenomena of
  ferromagnetic transition in double-exchange systems},\ }\href
  {https://doi.org/10.1143/JPSJ.70.1487} {\bibfield  {journal} {\bibinfo
  {journal} {Journal of the Physical Society of Japan}\ }\textbf {\bibinfo
  {volume} {70}},\ \bibinfo {pages} {1487} (\bibinfo {year}
  {2001})}\BibitemShut {NoStop}%
\bibitem [{\citenamefont {Ruderman}\ and\ \citenamefont
  {Kittel}(1954)}]{Ruderman1954}%
  \BibitemOpen
  \bibfield  {author} {\bibinfo {author} {\bibfnamefont {M.~A.}\ \bibnamefont
  {Ruderman}}\ and\ \bibinfo {author} {\bibfnamefont {C.}~\bibnamefont
  {Kittel}},\ }\bibfield  {title} {\bibinfo {title} {{Indirect Exchange
  Coupling of Nuclear Magnetic Moments by Conduction Electrons}},\ }\href
  {https://doi.org/10.1103/PhysRev.96.99} {\bibfield  {journal} {\bibinfo
  {journal} {Phys. Rev.}\ }\textbf {\bibinfo {volume} {96}},\ \bibinfo {pages}
  {99} (\bibinfo {year} {1954})}\BibitemShut {NoStop}%
\bibitem [{\citenamefont {Kasuya}(1956)}]{Kasuya1956}%
  \BibitemOpen
  \bibfield  {author} {\bibinfo {author} {\bibfnamefont {T.}~\bibnamefont
  {Kasuya}},\ }\bibfield  {title} {\bibinfo {title} {{A Theory of Metallic
  Ferro- and Antiferromagnetism on Zener's Model}},\ }\href
  {https://doi.org/10.1143/PTP.16.45} {\bibfield  {journal} {\bibinfo
  {journal} {Prog. Theor. Phys.}\ }\textbf {\bibinfo {volume} {16}},\ \bibinfo
  {pages} {45} (\bibinfo {year} {1956})}\BibitemShut {NoStop}%
\bibitem [{\citenamefont {Yosida}(1957)}]{Yosida1957}%
  \BibitemOpen
  \bibfield  {author} {\bibinfo {author} {\bibfnamefont {K.}~\bibnamefont
  {Yosida}},\ }\bibfield  {title} {\bibinfo {title} {Magnetic properties of
  {Cu-Mn} alloys},\ }\href {https://doi.org/10.1103/PhysRev.106.893} {\bibfield
   {journal} {\bibinfo  {journal} {Phys. Rev.}\ }\textbf {\bibinfo {volume}
  {106}},\ \bibinfo {pages} {893} (\bibinfo {year} {1957})}\BibitemShut
  {NoStop}%
\bibitem [{\citenamefont {Wei\ss{}e}\ \emph {et~al.}(2006)\citenamefont
  {Wei\ss{}e}, \citenamefont {Wellein}, \citenamefont {Alvermann},\ and\
  \citenamefont {Fehske}}]{weisse06}%
  \BibitemOpen
  \bibfield  {author} {\bibinfo {author} {\bibfnamefont {A.}~\bibnamefont
  {Wei\ss{}e}}, \bibinfo {author} {\bibfnamefont {G.}~\bibnamefont {Wellein}},
  \bibinfo {author} {\bibfnamefont {A.}~\bibnamefont {Alvermann}},\ and\
  \bibinfo {author} {\bibfnamefont {H.}~\bibnamefont {Fehske}},\ }\bibfield
  {title} {\bibinfo {title} {The kernel polynomial method},\ }\href
  {https://doi.org/10.1103/RevModPhys.78.275} {\bibfield  {journal} {\bibinfo
  {journal} {Rev. Mod. Phys.}\ }\textbf {\bibinfo {volume} {78}},\ \bibinfo
  {pages} {275} (\bibinfo {year} {2006})}\BibitemShut {NoStop}%
\bibitem [{\citenamefont {Barros}\ and\ \citenamefont {Kato}(2013)}]{barros13}%
  \BibitemOpen
  \bibfield  {author} {\bibinfo {author} {\bibfnamefont {K.}~\bibnamefont
  {Barros}}\ and\ \bibinfo {author} {\bibfnamefont {Y.}~\bibnamefont {Kato}},\
  }\bibfield  {title} {\bibinfo {title} {Efficient langevin simulation of
  coupled classical fields and fermions},\ }\href
  {https://doi.org/10.1103/PhysRevB.88.235101} {\bibfield  {journal} {\bibinfo
  {journal} {Phys. Rev. B}\ }\textbf {\bibinfo {volume} {88}},\ \bibinfo
  {pages} {235101} (\bibinfo {year} {2013})}\BibitemShut {NoStop}%
\bibitem [{\citenamefont {Wang}\ \emph {et~al.}(2018)\citenamefont {Wang},
  \citenamefont {Chern}, \citenamefont {Batista},\ and\ \citenamefont
  {Barros}}]{wang18}%
  \BibitemOpen
  \bibfield  {author} {\bibinfo {author} {\bibfnamefont {Z.}~\bibnamefont
  {Wang}}, \bibinfo {author} {\bibfnamefont {G.-W.}\ \bibnamefont {Chern}},
  \bibinfo {author} {\bibfnamefont {C.~D.}\ \bibnamefont {Batista}},\ and\
  \bibinfo {author} {\bibfnamefont {K.}~\bibnamefont {Barros}},\ }\bibfield
  {title} {\bibinfo {title} {{Gradient-based stochastic estimation of the
  density matrix}},\ }\href {https://doi.org/10.1063/1.5017741} {\bibfield
  {journal} {\bibinfo  {journal} {The Journal of Chemical Physics}\ }\textbf
  {\bibinfo {volume} {148}},\ \bibinfo {pages} {094107} (\bibinfo {year}
  {2018})}\BibitemShut {NoStop}%
\bibitem [{\citenamefont {Goodfellow}\ \emph {et~al.}(2016)\citenamefont
  {Goodfellow}, \citenamefont {Bengio},\ and\ \citenamefont
  {Courville}}]{Goodfellow16-ch9}%
  \BibitemOpen
  \bibfield  {author} {\bibinfo {author} {\bibfnamefont {I.}~\bibnamefont
  {Goodfellow}}, \bibinfo {author} {\bibfnamefont {Y.}~\bibnamefont {Bengio}},\
  and\ \bibinfo {author} {\bibfnamefont {A.}~\bibnamefont {Courville}},\
  }\bibfield  {title} {\bibinfo {title} {Convolutional networks},\ }in\ \href
  {http://www.deeplearningbook.org} {\emph {\bibinfo {booktitle} {Deep
  Learning}}}\ (\bibinfo  {publisher} {MIT Press},\ \bibinfo {year} {2016})\
  Chap.~\bibinfo {chapter} {9}, pp.\ \bibinfo {pages} {326--366}\BibitemShut
  {NoStop}%
\bibitem [{\citenamefont {Landau}\ and\ \citenamefont
  {Binder}(2014)}]{Landau_book14}%
  \BibitemOpen
  \bibfield  {author} {\bibinfo {author} {\bibfnamefont {D.~P.}\ \bibnamefont
  {Landau}}\ and\ \bibinfo {author} {\bibfnamefont {K.}~\bibnamefont
  {Binder}},\ }\href@noop {} {\emph {\bibinfo {title} {A Guide to Monte Carlo
  Simulations in Statistical Physics}}},\ \bibinfo {edition} {4th}\ ed.\
  (\bibinfo  {publisher} {Cambridge University Press},\ \bibinfo {year}
  {2014})\BibitemShut {NoStop}%
\bibitem [{\citenamefont {Glauber}(1963)}]{Glauber63}%
  \BibitemOpen
  \bibfield  {author} {\bibinfo {author} {\bibfnamefont {R.~J.}\ \bibnamefont
  {Glauber}},\ }\bibfield  {title} {\bibinfo {title} {{Time-Dependent
  Statistics of the Ising Model}},\ }\href {https://doi.org/10.1063/1.1703954}
  {\bibfield  {journal} {\bibinfo  {journal} {Journal of Mathematical Physics}\
  }\textbf {\bibinfo {volume} {4}},\ \bibinfo {pages} {294} (\bibinfo {year}
  {1963})}\BibitemShut {NoStop}%
\bibitem [{\citenamefont {Bray}(1994)}]{Bray1994}%
  \BibitemOpen
  \bibfield  {author} {\bibinfo {author} {\bibfnamefont {A.~J.}\ \bibnamefont
  {Bray}},\ }\bibfield  {title} {\bibinfo {title} {Theory of phase-ordering
  kinetics},\ }\href {https://doi.org/10.1080/00018739400101505} {\bibfield
  {journal} {\bibinfo  {journal} {Advances in Physics}\ }\textbf {\bibinfo
  {volume} {43}},\ \bibinfo {pages} {357} (\bibinfo {year} {1994})}\BibitemShut
  {NoStop}%
\bibitem [{\citenamefont {Onuki}(2002)}]{Onuki2002}%
  \BibitemOpen
  \bibfield  {author} {\bibinfo {author} {\bibfnamefont {A.}~\bibnamefont
  {Onuki}},\ }\href@noop {} {\emph {\bibinfo {title} {Phase Transition
  Dynamics}}}\ (\bibinfo  {publisher} {Cambridge University Press},\ \bibinfo
  {year} {2002})\BibitemShut {NoStop}%
\bibitem [{\citenamefont {Puri}\ and\ \citenamefont
  {Wadhawan}(2009)}]{Puri2009}%
  \BibitemOpen
  \bibfield  {author} {\bibinfo {author} {\bibfnamefont {S.}~\bibnamefont
  {Puri}}\ and\ \bibinfo {author} {\bibfnamefont {V.}~\bibnamefont
  {Wadhawan}},\ }\href@noop {} {\emph {\bibinfo {title} {Kinetics of phase
  transitions}}}\ (\bibinfo  {publisher} {CRC press},\ \bibinfo {year}
  {2009})\BibitemShut {NoStop}%
\bibitem [{\citenamefont {Allen}\ and\ \citenamefont {Cahn}(1972)}]{Allen1972}%
  \BibitemOpen
  \bibfield  {author} {\bibinfo {author} {\bibfnamefont {S.}~\bibnamefont
  {Allen}}\ and\ \bibinfo {author} {\bibfnamefont {J.}~\bibnamefont {Cahn}},\
  }\bibfield  {title} {\bibinfo {title} {Ground state structures in ordered
  binary alloys with second neighbor interactions},\ }\href
  {https://doi.org/https://doi.org/10.1016/0001-6160(72)90037-5} {\bibfield
  {journal} {\bibinfo  {journal} {Acta Metallurgica}\ }\textbf {\bibinfo
  {volume} {20}},\ \bibinfo {pages} {423} (\bibinfo {year} {1972})}\BibitemShut
  {NoStop}%
\bibitem [{\citenamefont {Corberi}(2015)}]{corberi15}%
  \BibitemOpen
  \bibfield  {author} {\bibinfo {author} {\bibfnamefont {F.}~\bibnamefont
  {Corberi}},\ }\bibfield  {title} {\bibinfo {title} {Coarsening in
  inhomogeneous systems},\ }\href
  {https://doi.org/https://doi.org/10.1016/j.crhy.2015.03.019} {\bibfield
  {journal} {\bibinfo  {journal} {Comptes Rendus Physique}\ }\textbf {\bibinfo
  {volume} {16}},\ \bibinfo {pages} {332} (\bibinfo {year} {2015})},\ \bibinfo
  {note} {coarsening dynamics / Dynamique de coarsening}\BibitemShut {NoStop}%
\bibitem [{\citenamefont {Shore}\ \emph {et~al.}(1992)\citenamefont {Shore},
  \citenamefont {Holzer},\ and\ \citenamefont {Sethna}}]{shore92}%
  \BibitemOpen
  \bibfield  {author} {\bibinfo {author} {\bibfnamefont {J.~D.}\ \bibnamefont
  {Shore}}, \bibinfo {author} {\bibfnamefont {M.}~\bibnamefont {Holzer}},\ and\
  \bibinfo {author} {\bibfnamefont {J.~P.}\ \bibnamefont {Sethna}},\ }\bibfield
   {title} {\bibinfo {title} {Logarithmically slow domain growth in nonrandomly
  frustrated systems: Ising models with competing interactions},\ }\href
  {https://doi.org/10.1103/PhysRevB.46.11376} {\bibfield  {journal} {\bibinfo
  {journal} {Phys. Rev. B}\ }\textbf {\bibinfo {volume} {46}},\ \bibinfo
  {pages} {11376} (\bibinfo {year} {1992})}\BibitemShut {NoStop}%
\bibitem [{\citenamefont {Evans}(2002)}]{evans02}%
  \BibitemOpen
  \bibfield  {author} {\bibinfo {author} {\bibfnamefont {M.~R.}\ \bibnamefont
  {Evans}},\ }\bibfield  {title} {\bibinfo {title} {Anomalous coarsening and
  glassy dynamics},\ }\href {https://doi.org/10.1088/0953-8984/14/7/302}
  {\bibfield  {journal} {\bibinfo  {journal} {Journal of Physics: Condensed
  Matter}\ }\textbf {\bibinfo {volume} {14}},\ \bibinfo {pages} {1397}
  (\bibinfo {year} {2002})}\BibitemShut {NoStop}%
\bibitem [{\citenamefont {Tanaka}(2000)}]{tanaka00}%
  \BibitemOpen
  \bibfield  {author} {\bibinfo {author} {\bibfnamefont {H.}~\bibnamefont
  {Tanaka}},\ }\bibfield  {title} {\bibinfo {title} {Viscoelastic phase
  separation},\ }\href {https://doi.org/10.1088/0953-8984/12/15/201} {\bibfield
   {journal} {\bibinfo  {journal} {Journal of Physics: Condensed Matter}\
  }\textbf {\bibinfo {volume} {12}},\ \bibinfo {pages} {R207} (\bibinfo {year}
  {2000})}\BibitemShut {NoStop}%
\end{thebibliography}%

\end{document}